\documentclass[journal=jctcce,manuscript=article,email=false,layout=twocolumn]{achemso}
\usepackage[utf8]{inputenc}
\usepackage{lipsum}
\usepackage{amsmath}
\usepackage{hyperref}
\usepackage{cleveref}
\usepackage{amsfonts}
\usepackage{amssymb}
\usepackage{bm}
\usepackage{lmodern}
\usepackage{graphicx}
\usepackage{booktabs}
\usepackage{subfig}
\usepackage[]{tabularx}
\usepackage{siunitx}
\usepackage{derivative}
\usepackage{tikz}
\usepackage{adjustbox}

\usetikzlibrary {3d, perspective}
\newcommand{\FE}{\mathcal{F}}

\newcommand*\conj[1]{\overline{#1}}
\DeclareMathOperator{\Root}{RootFind}
\DeclareMathOperator{\MSE}{MSE}

\graphicspath{{./figures/}{../figures/asy/}}

\newcommand{\hbm}[1]{
  \hat{\bm{#1}}
}
\newcommand{\tj}[6]{
\begin{pmatrix}
  #1 & #2 & #3 \\
  #4 & #5 & #6
\end{pmatrix}}

\newcommand{\wD}[0]{D}
\newcommand{\pOp}[0]{\Pi}

\date{}

\title{Machine learning of a density functional for anisotropic patchy particles }
\author{Alessandro Simon}
\affiliation{Institute for Applied Physics, University of Tübingen, Auf der Morgenstelle 10, 72076 Tübingen, Germany}
\alsoaffiliation{Max Planck Institute for Intelligent Systems, Max-Planck-Ring 4, 72076 Tübingen, Germany}
\author{Jens Weimar}
\affiliation{Institute for Applied Physics, University of Tübingen, Auf der Morgenstelle 10, 72076 Tübingen, Germany}
\author{Georg Martius}
\affiliation{Max Planck Institute for Intelligent Systems, Max-Planck-Ring 4, 72076 Tübingen, Germany}
\author{Martin Oettel}
\affiliation{Institute for Applied Physics, University of Tübingen, Auf der Morgenstelle 10, 72076 Tübingen, Germany}
\begin{document}
\begin{abstract}
    Anisotropic patchy particles have become an archetypical statistical model system for associating fluids. Here we formulate an approach to the Kern--Frenkel model via classical density functional theory to describe the positionally and orientationally resolved equilibrium density distributions in flat wall geometries. The density functional is split into a reference part for the orientationally averaged density and an orientational part in mean-field approximation. To bring the orientational part into a kernel form suitable for machine learning techniques, an expansion into orientational invariants and the proper incorporation of single-particle symmetries is formulated. The mean-field kernel is constructed via machine learning on the basis of hard wall simulation data. {Results are compared to the well-known random-phase approximation which strongly understimates the orientational correlations close to the wall.}
    Successes and shortcomings of the mean-field treatment of the orientational part are highlighted and perspectives are given for attaining a full density functional via machine learning.  
\end{abstract}
\maketitle
\section{Introduction}

A useful model system for describing various phenomena in soft matter physics is given by patchy particles where particles with an isotropic repulsive core additionally interact via a certain number of attractive bonding sites distributed over their surface. These patchy particle systems show novel phenomena such as the existence of empty liquids \cite{bianchi_phase_2006, heras_phase_2011} or of stable equilibrium gels \cite{smallenburg_liquids_2013, sciortino_equilibrium_2017}. The Kern-Frenkel (KF) model is an example for such a patchy particle model with a mathematical form for the pair potential that is easy to simulate \cite{kern_frenkel2003}. In fact, many theoretical insights, especially on phase behavior, have been obtained by simulations \cite{foffi2007possibility,romano2010phase}. An alternative approach to simulations for obtaining equilibrium properties of soft matter model systems is (classical) density functional theory (DFT). It has been frequently and successfully employed for isotropic fluids \cite{lutsko2010recent,evans2009density}, in particular for hard particles in the form of the very precise fundamental measure theory (FMT), for a review see Ref.~\cite{roth2010fundamental}. For the specific case of patchy particles, FMT-based functionals have been derived \cite{yuwu2002,stopper2018bulk} which are functionals of the \textit{orientationally averaged} density profile. These functionals describe various properties of patchy particles reasonably well (e.g. phase diagrams and orientationally averaged pair correlations), but there are limitations which are clearly due to the neglect of orientational correlations (e.g. density profiles around hard tracer particles). Thus, it is desirable to find free energy functionals of the full orientation-dependent density profile. However, there are no analytical methods known to construct such functionals for anisotropic particles that are similar in accuracy as FMT for hard particles. Hence, it seems to be promising to turn to numerical and data-driven methods such as machine learning (ML) for this task.               

Finding classical density functionals with ML methods is a rather recent development. Initial work has focused on model fluids in one dimension (1D) \cite{shang2019classical,shang2020FEQL,yatsyshin2021,wu2022}, whereas work on more realistic systems in 3D is scarce, for a study of the 3D Lennard-Jones (LJ) fluid at one supercritical temperature, see Ref.~\cite{cats2021} and for a study of hard spheres in planar geometry see the very recent Ref.~\cite{sammuller2023neural}. Among these works, one can distinguish between approaches to learning integral kernels for the generally unknown, but sensibly parametrized part in the functional describing attractive interactions \cite{shang2019classical,cats2021}, approaches to learn an analytic expression \cite{shang2020FEQL} or numerical representations for the entire functional \cite{sammuller2023neural} and studies more focussing on the uncertainty assessment of the learned density functional maps \cite{yatsyshin2021,wu2022}. The ML literature on density functionals for the quantum electron problem goes back a few more years \cite{burke2012}, recent developments show interesting parallels to the approaches taken for classical systems, as e.g. a work on finding analytic functionals in \cite{lili2022} or the use of the minimizing equations in the ML networks \cite{lili2021}. 

In this paper, we aim at finding a density functional for the KF model with ML methods, describing the orientational correlations from Monte Carlo simulation data between hard walls. The motivation is also to test these ML methods for a complex 3D system where there is little previous experience with analytical DFT. The approach is similar in spirit to Refs.~\cite{shang2019classical,cats2021} in that an \textit{ansatz} for the unknown part in the functional is chosen which employs weighted densities and ML is used to determine these weights. We restrict our \textit{ansatz} to a mean-field form where the learned weights have a straightforward interpretation and may be compared to standard forms from liquid state theory. We note that mean-field forms have been employed before in DFT studies of anisotropic particles; see Refs.~\cite{cattes2016,wandrei2018mean} for a Heisenberg-type fluid where the mean-field form is directly taken from the anisotropic, attractive part of the interaction potential (random-phase approximation or RPA), or Ref.~\cite{teixeira2019patchy} for a patchy particle fluid with two patches where the orientational moments in the mean-field part are assumed to be of Lennard-Jones type and free parameters are fitted. In contrast, the present study introduces a sensible method for obtaining an ``optimized'' mean-field form for the orientational part of the density functional, constrained by simulation data. For anisotropic fluids, even simple mean-field functionals are technically challenging: the density depends on six variables, three spatial coordinates and three orientation angles. A systematic expansion of the orientational part proceeds via Wigner D-matrices and orientational invariants, pioneered in integral equation studies of anisotropic fluids \cite{blum1972invariant} and developed more recently in the context of integral equation and DFT studies of water in Refs.~\cite{ding2017efficient,belloni2017exact,jeanmairet2013molecular,borgis2021accurate}. 
An earlier work where anisotropic mean-field DFT was used to model a site-site model of water, including the liquid-vapor interface is Ref.~\cite{yang1994density}
This expansion also allows controlled truncations. Nevertheless, to make the problem suitable for machine learning, some basic formalism regarding the orientational expansion and the proper incorporation of single-particle (molecular) symmetries needs to be laid out, which (besides the actual machine learning of the mean-field functional) is one of the main aims of the present paper.    
We hope that the method can be extended in the future to tackle the task of finding the ``full'' functional for patchy and other anisotropic particles.     

The paper is structured as follows: First, we specify the anisotropic two-body potential and its consequences for the density distribution due to tetrahedral symmetry (Sec.~\ref{sec:kf}--\ref{sec:pair}). In Sec.~\ref{sec:dft} the density functional formalism is introduced, together with a short description of functionals that have been used to up to this point to model the structure of the fluid. This is followed by the introduction of the mean-field expansion to the excess free energy functional for the  angular degrees of freedom.
In Sec.~\ref{sec:kfsim} we present simulation results of the Kern-Frenkel system, consisting of density profiles for the confined fluid between two walls and the angular distribution of the orientation of the constituent particles. A numerically stable machine learning procedure for fitting mean-field kernels to the simulation data is proposed (Sec.~\ref{sec:ml}) and %
results are presented for a range of supercritical temperatures (Sec.~\ref{sec:resultsML}). In Sec.~\ref{sec:summary}. we conclude with a summary and an outlook on future perspectives.

\section{Theory}
\subsection{Kern-Frenkel potential}
\label{sec:kf}

The KF fluid we consider is modeled by hard spheres possessing four attractive patches placed at the corners of a tetrahedron which is centered in the middle of the particle. 
Besides the tetrahedron condition (angle between patch vectors equal to  $\arccos ( -1/3)$) one is still free to chose the exact placement of the patches in the molecule's body-fixed reference frame. Depending on this convention, the symmetry conditions 
in terms of the Euler angles change, which can be useful later on. Two popular choices are the following:

\paragraph*{Convention A}

Here the patches are placed at the following positions
\begin{equation}
P_A = \frac{1}{\sqrt{3}}\begin{pmatrix}
  -1& 1 & 1 & -1\\
  -1 & 1 & -1 & 1\\
  1 & 1 & -1 & -1
\end{pmatrix}  
\end{equation}
where each column $\alpha$ is a unit vector $\hat{\bm{r}}^{\alpha}$ pointing from the center to the patch. All vector components have
the same magnitude, and it is rather easy to see which mirror symmetries the molecule possesses.

\paragraph*{Convention B}

We rotate the configuration from the previous convention by the angle $\frac{\pi}{4}$ (actively) around the $z$ axis to get convention (B). Here, a pair of patches
is placed along the $xz$ plane, while the other pair sits on the $yz$ plane.
The positions are
\begin{equation}
  P_B = \frac{1}{\sqrt{3}}\begin{pmatrix}
    -\sqrt{2} & \sqrt{2} & 0 & 0\\
    0 & 0 & -\sqrt{2} & \sqrt{2}\\
    1 & 1 & -1 & -1
  \end{pmatrix}  
  \end{equation}
The reason we prefer this convention is that two symmetry planes fall together with the coordinate planes. 
In principle, however, any choice is permissible.

\begin{figure}%
  \centering
  \subfloat[\centering Convention (A)]{{\includegraphics[width=0.7\columnwidth]{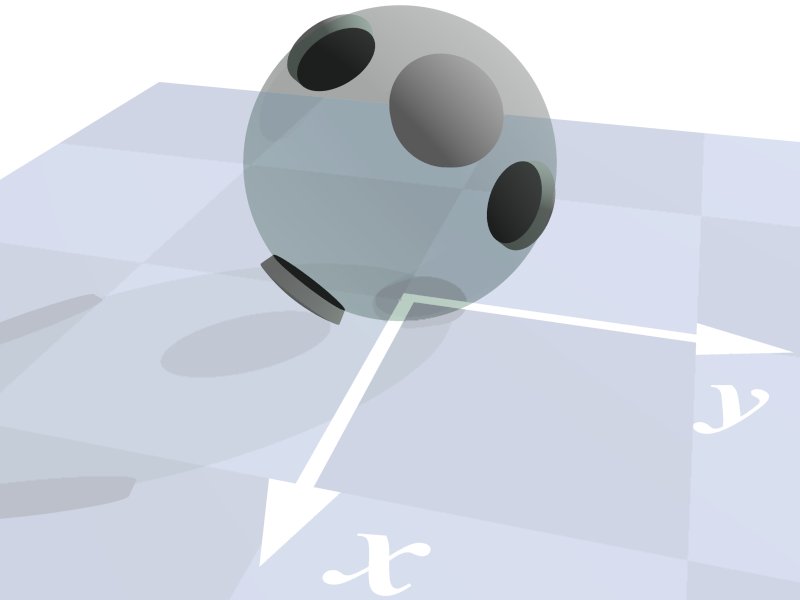} }}%
  \qquad \qquad
  \subfloat[\centering Convention (B)]{{\includegraphics[width=0.7\columnwidth]{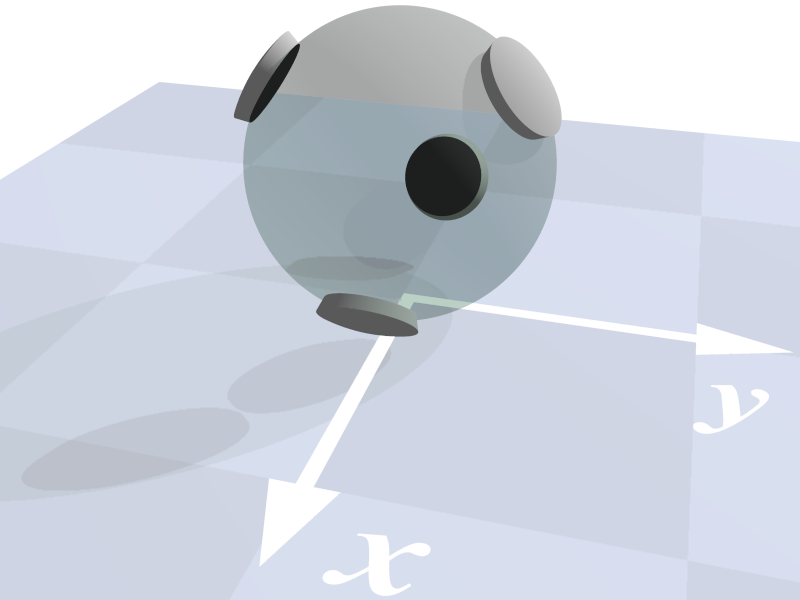} }}%
  \caption{Two possible conventions that are used for the placement of the patches in the body-fixed system. Note that convention (A) had been used in~\cite{steele1980symmetry} and it is the standard convention in the simulation code described in~\cite{rovigatti2018simulate}.}
  \label{fig:example}%
\end{figure}

{The KF potential between a pair of molecules is a sum of a hard sphere part $\phi_\text{HS}(r)$ for diameter $\sigma$ and
an anisotropic patch part $V^\text{KF}$ that depends on the individual orientation of the molecules} and their
positions relative to each other. For parametrizing the orientation of a three dimensional body
we will use the well known Euler angle parametrization with the angles
$(\phi, \theta, \chi) \equiv \Omega$ and 
\begin{equation}
 \int d\Omega:= \int_0^{2\pi} \odif{\phi} \int_0^\pi \odif{\theta} \int_0^{2\pi} \odif{\chi} = 8 \pi^{2} =: K.    
\end{equation}

The anisotropic part of the potential is defined by
\begin{equation}
  \label{eq:16}
  \footnotesize V^\mathrm{KF}(\bm{r}_{12}, \Omega_{1}, \Omega_{2}) =  \phi_{\mathrm{sw}}(r_{12})\!\!\sum_{\alpha,\beta=1}^{N_\text{p}}\!\!  \phi_{\text{p}}(\bm{r}_{12}, \hat{\bm{r}}_{1}^{\alpha}(\Omega_{1}), \hat{\bm{r}}_{2}^{\beta}(\Omega_{2}))
\end{equation}
Here, $N_\text{p}$ is the number of patches and   $\phi_{\text{sw}}(r)$ is a square well potential
\begin{equation}
  \phi_{\text{sw}}(r)~=~ \left\{ \begin{matrix} -\epsilon \qquad (r \in [\sigma,\sigma+\delta]) 
      \\ 0 \qquad \text{(otherwise)} \end{matrix} \right.
\end{equation}
with the attractive range $\delta$ and attraction depth $\epsilon$. $\phi_{\text{p}}$ is the orientational patch-patch part
\begin{equation}
  \label{eq:17} \footnotesize
  \phi_{\text{p}}(\bm{r}_{12}, \hat{\bm{r}}_{1}^{\alpha}(\Omega_{1}), \hat{\bm{r}}_{2}^{\beta}(\Omega_{2}))  = \begin{cases}
    1 &\text{if}\begin{cases}
      \phantom{-}\hat{\bm{r}}_{12} \cdot \hat{\bm{r}}_{1}^{\alpha} > \cos \theta_{\text{max}}\\
      -\hat{\bm{r}}_{12} \cdot \hat{\bm{r}}_{2}^{\beta} > \cos \theta_{\text{max}}
    \end{cases}\\
    0 &\text{else}
    \end{cases}
  \end{equation}
characterized by the opening angle $\theta_{\text{max}}$ of the ``cones'' that interact attractively between the particles. 
{Here, $\bm{r}_{12}=\bm{r}_{1}-\bm{r}_{2}$ is the vector connecting the two centers of the particles, and $\hat{\bm{r}}_i^{\alpha}$ is a unit vector from the  center  of  particle $i$ to  a  patch $\alpha$ on  its  surface  depending  on  the  particle orientation $\Omega_i$.}
In general, depending on the chosen parameters,
more than two patches could be bonded. For certain choices of $\delta$ and $\theta_\text{max}$ 
this is avoided (single bond condition) and this will be adopted in the actual calculations, see below.

\subsection{Expansion of the density distribution}

Owing to the orientational dependence of the density distribution on three angles, its behavior could be rather complex and visualizations are not easy. The complexity is reduced by expanding the 
orientational part in a suitable functional basis, the Wigner D-matrices $\wD^l_{mn}(\Omega)$ \cite{rose1995elementary, gray84theory}
These are given by
\begin{equation}
  \label{eq:15}
  \wD^{l}_{mn}(\phi, \theta, \chi) = e^{-i m \phi} d^{l}_{mn}(\theta) e^{-i n \chi}
\end{equation}
with the integer indices $m,n$ running from $-l$ to $l$ and $l \in \mathbb{N}$.
Then, the forward and backward transformations of a sufficiently well-behaved function $f$ that may also depend on other variables, here $\bm{r}$, are given by
\begin{gather}
  f(\bm{r}, \Omega) = \sum_{lmn} f^{l}_{mn}(\bm{r}) \wD^{l}_{mn}(\Omega) \label{eq:d_expansion}\\
  f^{l}_{mn}(\bm{r}) = \frac{2l + 1}{K} \int \odif{\Omega} f(\bm{r}, \Omega) \conj{\wD^{l}_{mn}}(\Omega)  \label{eq:expansion_coeff}\;,
\end{gather}
{where $\conj{\wD^{l}_{mn}}$ is the complex conjugate of $\wD^{l}_{mn}$.}
We call $f^{l}_{mn}$ the (orientational) moments of the function $f$.%

We adopt {Euler angles in the $zyz$ prescription to describe the rotations of the molecule.} Explicitly, the body-fixed reference frame is first rotated by the 
angle $\phi$ around the original $z$ axis, resulting in two new perpendicular axes $x'$ and $y'$. Then we rotate the frame by the angle $\theta$ around
$y'$, again resulting in two new axes $x''$ and $z''$, while $y'' \equiv y'$. The final rotation of $\chi$ is done around $z''$. The order of these operations is also 
reflected in the arguments of the D-matrix. Alternatively, the same rotation can be achieved by rotating around space-fixed axes $XYZ$ with the
same angles but in reverse order, i.e.\ by $\chi$ around $Z$, then by $\theta$ around $Y$ and then finally by $\phi$ around $Z$ again.

\subsubsection{Symmetry constraints}

The existence of symmetries leads to certain moments in the above expansion to vanish or to become linearly dependent.
Two kinds of symmetries arise in our context. Firstly, due to the symmetry of the molecule and the indistinguishability of the patches, a certain rotational configuration
$\Omega$ is physically equivalent to another configuration $\Omega' \neq \Omega$. This is the molecular symmetry, in our case the (chiral or rotational)
tetrahedral group $T$, containing 12 elements. The full tetrahedral $T_d$ group also includes reflections and contains 24 elements in total.
The groups $T$ and $T_d$ are isomorphic to the alternating group $A_4$ and the symmetric group $S_4$, respectively. Thus, it is possible to 
label the symmetry rotations of the tetrahedral group with the permutation of the four corners of the tetrahedron.

Secondly, the external potential may also introduce additional symmetry constraints on the density and orientation distribution. In our case of an orientation-independent flat wall potential, this will be an additional rotational symmetry perpendicular to a wall normal; see below.

For treating the symmetry of the molecules, we adapt the exposition in ref.~\cite{steele1980symmetry} to our purpose.
Suppose that the function $f(\Omega)$ is invariant under the group $G$, that is
\begin{equation}
  \label{eq:invariance_f}
  f(\Omega) = f(\Omega') = f(\Omega'') = \cdots
\end{equation}
where the primed orientations correspond  to orientations ``congruent'' to the initial one and result from acting on $\Omega$ with one group element $g \in G$. This is equivalent to permuting patches before bringing the molecule to orientation $\Omega$.
Each of the previous $f(\cdot)$ can be expanded into Wigner matrices as in ~\cref{eq:d_expansion}.  $D(\Omega')$ is related to $D(\Omega)$ through a transformation law for
Wigner matrices under rotation of the body-fixed axes, 
\begin{equation}
\wD^l_{kn}(\Omega') =  \wD^l_{kn}(\Omega R) = \sum_{m} \wD^l_{km}(\Omega)  \wD^l_{m n}(R),
\end{equation}
see, e.g., Ch. 4.7, eq. (1) in \cite{varshalovich1988quantum}, or eq. (A.101) in~\cite{gray84theory}. 

Note that here the Wigner matrix serves in principle two purposes. Once it is used as a basis for $f \in L^2(\mathrm{SO}(3))$
and a second time to rotate these basis functions.
We can now sum over all the terms in eq.~\eqref{eq:invariance_f} and divide by the number $|G|$ of symmetry elements  arriving at
\begin{equation}
  \label{eq:7}
  \frac{f(\Omega) + f(\Omega') + \cdots}{|G|} = f(\Omega)
\end{equation}
Expanding both sides in the Wigner basis leads to the following condition
\begin{multline}
  \label{eq:5}
  {\sum_{lmn}} f^l_{mn} \wD^l_{mn}(\Omega)  \\ =  {\sum_{lmn}} f^l_{mn} {\sum_k}\wD^l_{mk}(\Omega)  \frac{1}{|G|} \sum_{R_i \in G} \wD^l_{k n}(R_i) \\ 
  =
  {\sum_{lmn}} f^l_{mn} {\sum_k} \wD^l_{mk}(\Omega)  \pOp^{(l)}_{kn}
\end{multline}
where we define the projection operator as
\begin{equation}
  \label{eq:proj_op}
 \frac{1}{|G|}  \sum_{R_i \in G} \wD^l_{k n}(R_i) =   \pOp^{(l)}_{kn}
\end{equation}
In eq.~\eqref{eq:5}, one can project onto the moments $f^l_{mn}$ by applying $(2l+1)/K \int d\Omega \conj{D^{l'}_{m'n'}} \cdots $ to both sides. We obtain:
\begin{equation}
  \label{eq:8}
  f^l_{mn} = f^l_{mk} \pOp^{(l)}_{nk}
\end{equation}
which corresponds to a linear equation for every row $m$ of the coefficient matrix $f^l_{mn}$, written in the usual form as
\begin{equation}
  \label{eq:9}
  \pOp^{(l)}_{nk} f^l_{\circ k} = f^l_{\circ n}
\end{equation}
Solving this equation leads to the sought symmetry conditions of the expansion coefficients.

This is equivalent to the fact that not every Wigner matrix can appear in the expansion of an invariant function. Just like the function itself, the basis functions that appear must also be invariant and this can only be achieved by certain linear combinations of Wigner matrices. We can write the new, symmetric basis as
\begin{equation}
  \label{eq:27}
  \Delta^l_{m}(\Omega) = \sum_{n} d^l_n \wD^l_{mn}(\Omega)
\end{equation}
Repeating the above calculation for the invariant $\Delta^l_{m}(\Omega)$ instead of
the $f(\Omega)$ leads to the same linear equation, only this time with the interpretation of having constructed a $G$ invariant
basis of $L^2(\mathrm{SO}(3))$. The resulting eigenvalue problem is the same as in \cref{eq:9}
\begin{equation}
  \label{eq:30}
\sum_k    \pOp^{(l)}_{nk} d^l_{k} = d^l_{n}
\end{equation}
In principle there could be $2l+1$ orthogonal eigenvectors for every value of $l$ (imagine the case without any symmetry) but usually the number of
eigenvectors with eigenvalue $1$ is much smaller. We will therefore label the solutions by an index $j$
as in 
\begin{equation}
  d^l_{[j]n}
\end{equation}
such that we can also label the basis functions accordingly
\begin{equation}
  \Delta^l_{m[j]}(\Omega) = \sum_{n} d^l_{[j]n} \wD^l_{mn}(\Omega)
\end{equation}
In order for the new basis function to fulfill the same normalization as the Wigner matrices one needs to scale the $d^l_{[j]}(n)$ such that
\begin{equation}
\sum_n (d^l_{[j]n})^2 = 1
\end{equation}

\begin{table}
\begin{center}
  \adjustbox{width=\columnwidth}{
  \begin{tabular}{@{}c@{}ccc|ccc@{}} 
  \toprule
  permutation & $\phi$ & $\theta$ & $\chi$ &$\phi$&$\theta$&$\chi$  \\
  \midrule
  (id) & 0 & 0 & 0 & 0&0&0\\
  (012) & 0&$\pi/$2&$-\pi/2$ &$ -\pi/4$&$ \pi/2$& $-\pi/4$\\
  (031) & $\pi$& $\pi/2$&$\pi/2$ &$3\pi/4$&$ \pi/2$&$ 3\pi/4$ \\ 
  (132) & $\pi$& $\pi/2$&$-\pi/2$ &$ 3\pi/4$& $\pi/2$&$ -\pi/4$\\ 
  (023) & 0& $\pi/2$&$\pi/2$ &$-\pi/4$& $\pi/2$&$ 3\pi/4$\\  

  (021) &$-\pi/2$ & $ \pi/2$ & $\pi $& $-3\pi/4$ & $ \pi/2$ & $ -3\pi/4$\\ 
  (013) &$ \pi/2$ & $\pi/2$ &0 &$ \pi/4$& $\pi/2$&$ \pi/4$\\ 
  (123) &$ -\pi/2$& $\pi/2$&0 &$ -3\pi/4$&$ \pi/2$&$ \pi/4$\\ 
  (032) &$  \pi/2$& $\pi/2$&$\pi$ &  $\pi/4$& $\pi/2$&$ -3\pi/4$\\
    
  (02)(13) & 0&$\pi$&0 &$ -\pi/2$& $\pi$& 0\\
  (01)(23) & $\pi$&0&0 & $-\pi$& 0& 0\\ 
  (03)(12) & $\pi$&$\pi$&0 &$\pi/2$& $\pi$& 0 \\
    \midrule
convention & & (A) & & & (B) & \\
  \bottomrule
  \end{tabular}
  }
  \caption{\label{tab:angles_T}Angles for the rotations in the group $T$ for two different conventions of the body frame }
  \end{center}
\end{table}

\begin{table}
\centering
\begin{tabular}{ccc}
\toprule
  $l$ & non-zero & independent \\
    & $D^l_{mn}$ & $\Delta^l_{m[j]}$ \\\midrule
  0 & 0 & 0 \\
  1 & 0 & 0 \\
  2 & 0 & 0 \\
  3 & 2 & 1 \\
  4 & 3 & 1 \\
  5 & 0 & 0 \\
  6 & 7 & 2 \\
  7 & 4 & 1 \\
  8 & 5 & 1 \\
  9 & 8 & 2 \\\bottomrule
\end{tabular}
\caption{The number of non-vanishing $D^l_{mn}$ and linearly independent $\Delta^l_{m[j]}$ up to $l=9$, resulting from tetrahedral symmetry, for a fixed $m$. For a general inhomogeneous situation, $m=-l...l$, whereas in the flat wall geometry
(cylindrical symmetry), $m=0$.}
\label{tab:number_D}
\end{table}

\begin{table}
\adjustbox{width=\columnwidth}{
\begin{tabular}{@{}cc@{}c@{}}
\toprule
  $l$ & basis& components \\
      & function & \\\midrule
  3 & $\Delta^3_{m[1]}$ & $ \frac{1}{\sqrt 2} \wD^3_{m 2} + \frac{1}{\sqrt 2} \wD^3_{m \underline{2}}$ \\
  4 & $\Delta^4_{m[1]}$ & $ \frac{\sqrt 30}{12}\wD^4_{m \underline4} + \frac{\sqrt 30}{12} \wD^4_{m 4} -\frac{\sqrt 21}{6} \wD^4_{m 0}$ \\
  6 & $\Delta^6_{m[1]}$ & $ \frac{\sqrt 7}{4} \wD^6_{m \underline{4}} + \frac{\sqrt 7}{4} \wD^6_{m 4} + \frac{\sqrt 2}{4} \wD^6_{m 0} $  \\
      & $\Delta^6_{m[2]}$ & $- \frac{\sqrt{10}}{8} \wD^6_{m \underline{6}} + \frac{\sqrt{10}}{8} \wD^6_{m 6} -  \frac{\sqrt{22}}{8} \wD^6_{m\underline{2}} + \frac{\sqrt{22}}{8} \wD^6_{m 2} $ \\
  \bottomrule
\end{tabular}
}
\caption{Linear combinations of Wigner matrices that are invariant under all tetrahedral group transformations up to $l=6$
}
\label{tab:basis_fn}
\end{table}

\subsubsection{Tetrahedral symmetry}

After settling for one patch convention we can determine the Euler angles corresponding to every group element and then compute the "projection operator" 
from eq.~\eqref{eq:proj_op}. The necessary angles which enter the D-matrices are shown in \cref{tab:angles_T} for both patch conventions. Having chosen convention (B) we arrive at the following: $\pOp^{(1)} = \pOp^{(2)} = 0$, that means no Wigner matrices with $l=1, 2$ can appear in the expansion. Only starting with $l=3$ we start to see non-vanishing contributions to the expansion. Further, even though there are multiple non-vanishing moments for $l=\{3, 4\}$, in both cases they are linearly dependent, thus reducing the number of physically meaningful basis function to one.

The number of nonvanishing D-matrices as well as of the $\Delta$ invariants are shown in table~\ref{tab:number_D} normalized basis functions up to $l=6$ are shown in table~\ref{tab:basis_fn}

\subsubsection{Cylindrical symmetry (wall)}

Assuming cylindrical symmetry of the external potential perpendicular to the  $z$ axis (i.e. a flat wall) further reduces the number of non-vanishing moments. The external potential is now only $z$ dependent, hence
$\rho(\bm{r},\Omega)  \to \rho(z, \Omega)$. The density must now also be invariant under rotations of the molecule in the $xy$ plane (this leaves constant 
the distance of the individual patches with respect to the wall). These rotations correspond precisely to the final rotation (in reverse Euler order) with $\phi$ around the 
$Z$ axis. In other words we need
\begin{equation}
    \rho(z, \phi, \theta, \chi) = \rho(z, \phi + \delta , \theta, \chi) \qquad \forall \delta \in \left[0, 2\pi\right)
\end{equation}
which can only be achieved when $\rho$ does not depend on $\phi$. Hence only the moments $\rho^l_{0n}$ can be different from zero.

\subsection{Pair potential in rotational invariants}
\label{sec:pair}
The anisotropic  part of the Kern-Frenkel potential can be expanded in so called rotational invariants \cite{gray84theory,blum1972invariant,ding2017efficient}
\begin{equation}
  \label{eq:kf_ex}
  V^{\text{KF}}(\bm{r}, \Omega_{1}, \Omega_{2}) = \sum_{m n l \mu\nu} V^{mnl}_{\mu \nu}(r) \Phi^{mnl}_{\mu \nu}(\Omega_{1}, \Omega_{2}, \hat{\bm{r}})
\end{equation}
The basis functions $\Phi^{mnl}_{\mu \nu}$ are a linear combination of D-matrices such that the projected coefficients $V^{mnl}_{\mu \nu}$ are independent of the choice for the fixed reference frame, hence
the name rotational invariants. They are given by \cite{gray84theory}
\begin{multline}
  \label{eq:28}
  \Phi^{mnl}_{\mu \nu}(\Omega_{1}, \Omega_{2}, \hbm{r}) = c_{m}c_{n}\sum_{\mu' \nu' \lambda'} \tj{m}{n}{l}{\mu'}{\nu'}{\lambda'}  \\ \times \conj{\wD^{m}_{\mu' \mu}}(\Omega_{1}) \conj{\wD^{n}_{\nu' \nu}}(\Omega_{2}) \conj{\wD^{l}_{\lambda' 0}}(\hat{\bm{r}})
\end{multline}
with $c_{n} = \sqrt{2n+1}$ and the Wigner $3j$ symbol in round brackets which is a combinatorial quantity similar to Clebsch-Gordan coefficients but having more intuitive symmetry properties.
In the definition of the basis function, the orientations of the two particles, and their separation vector factorize into different Wigner matrices. This means that we can change the orientation of one particle to a symmetrically equivalent one, without affecting the other particle or their separation vector. It is therefore easy to see how the symmetry properties of a single particle reduce the number of possible coefficients in the expansion of the pair potential. We replace the Wigner matrix for each particle in eq.~\eqref{eq:28} with the corresponding symmetrized basis function, {eq.~\eqref{eq:27}}. %

We thus redefine $\Phi$ to the invariants
\begin{multline}
 \label{eq:Phi_Delta}
    \Phi^{mnl}_{[ji]}(\Omega_{1}, \Omega_{2}, \hbm{r}) = c_{m}c_{n}\sum_{\mu' \nu' \lambda'} \tj{m}{n}{l}{\mu'}{\nu'}{\lambda'}  \\ \times \conj{\Delta^{m}_{\mu' [j]}}(\Omega_{1}) \conj{\Delta^{n}_{\nu' [i]}}(\Omega_{2}) \conj{\wD^{l}_{\lambda' 0}}(\hat{\bm{r}})
\end{multline}

From the basis function derived before we see that the first non-vanishing pair potentials moments are
\begin{equation}
    \Phi^{000}_{[11]},
    \Phi^{033}_{[11]}, 
    \Phi^{330}_{[11]}, \Phi^{332}_{[11]}, \Phi^{334}_{[11]}, \ldots
\end{equation}
The fact that we are dealing with identical particles means that the moments resulting
from projections of permuted indices are equal up to factor dependent on $l$
\begin{equation}
V^{mnl}_{[ij]} = (-1)^l V^{nml}_{[ji]}
\end{equation}

\subsection{Orientational density functional theory}
\label{sec:dft}

\subsubsection{General formalism}
\label{sec:dft_general}
The main objective of density functional theory is finding an analytical expression for the grand potential functional
$\Xi[\rho]$ which is a functional of the density profile $\rho(\bm{r}, \Omega)$. The equilibrium state is subsequently found by minimization
\begin{equation}
  \label{eq:18}
  \fdv{\Xi[\rho]}{\rho(\bm{r}, \Omega)} = 0\,,\quad \text{for } \rho = \rho_{\mathrm{eq}}(\bm{r}, \Omega)
\end{equation}
The grand potential functional is decomposed into an intrinsic free energy functional $\FE$ and
a one-body term involving  the external potential $V^{\text{ext}}$ and the chemical potential $\mu$:
\begin{equation}
  \label{eq:19}\small
  \Xi = \FE[\rho(\bm{r}, \Omega)] + \int \odif{\bm{r}} \odif{\Omega}\, \rho(\bm{r}, \Omega) \left[ V^{\text{ext}}(\bm{r},\Omega) - \mu \right]
\end{equation}
One can further split the free energy into an ideal, non-interacting term $\FE_{\text{id}}$ and the excess term $\FE_{\text{ex}}$ that results
from particle interactions
\begin{equation}
  \label{eq:20}
  \FE[\rho(\bm{r}, \Omega)] = \FE_{\text{id}}[\rho(\bm{r}, \Omega)] + \FE_{\text{ex}}[\rho(\bm{r}, \Omega)] \;.
\end{equation}
The ideal gas term $\FE_{\text{id}}$ is given by
\begin{equation}
  \label{eq:21}
   \FE_{\text{id}} = k_{B}T \int \odif{\bm{r}} \odif{\Omega}\, \rho(\bm{r}, \Omega) \left[\ln(\rho \Lambda^{3} K) -1  \right]
 \end{equation}
 and the second, excess term is in general unknown. {Here, $\Lambda$ is the thermal de-Broglie length.}
In order to separate the free energy contributions due to the orientation of the particles from the {orientation-independent} effects we split $\rho(\bm{r}, \Omega)$
into {an orientationally averaged} component $\rho(\bm{r})$ (depending only on $\bm{r}$) and an orientational distribution $\alpha(\bm{r}, \Omega)$
\begin{equation}
  \label{eq:product_ansatz}
  \rho(\bm{r}, \Omega) = \rho(\bm{r}) \alpha(\bm{r}, \Omega) \;.
\end{equation}
{The orientational distribution is normalized:}
\begin{equation}
  \label{norm_alpha}
  {\int d\Omega  \alpha(\bm{r}, \Omega)=1.}
\end{equation} 

The ideal part of the Helmholtz free energy in the anisotropic case becomes
\begin{multline}
  \label{eq:6}
  \beta \FE_{\text{id}} = \int \odif{\bm{r}} \rho(\bm{r}) \left[ \log(\Lambda^{3}\rho(\bm{r})) - 1 \right] \\
  + \int \odif{\bm{r}}\odif{\Omega}\, {\rho(\bm{r})} \alpha(\bm{r},\Omega) \log(K \alpha(\bm{r},\Omega))
\end{multline}
{We propose to split the excess free energy into an isotropic and an orientational part,}
\begin{equation}
  \label{eq:22}
\FE_{\text{ex}} = \FE_{\text{ex,iso}}[c] +  \FE_{\text{ex,or}}[\rho(\bm{r}), \alpha(\bm{r}, \Omega)]
\end{equation}
thus expecting that some parts of the free energy are entirely determined by the orientationally
averaged fluid density $\rho(\bm{r})$ alone (such as the hard-core interaction), while others will probably also need
information about the orientations of the particles. Anisotropic interactions of which the angular dependence was integrated out also fit into the first category.

By minizing the grand canonical ensemble $\Xi = \FE + \int \rho (V^{\text{ext}} - \mu)$ w.r.t. the density $\rho(\bm{r})$ and the orientation distribution $\alpha(\bm{r}, \Omega)$
{ under the normalization constraint \eqref{norm_alpha}}
we arrive at the following Euler-Lagrange (EL) equations
\begin{multline}
\label{eq:2}
\rho(\bm{r}) = \exp \biggl( -\int \odif{\Omega}\, \alpha \log(K \alpha) + \beta \mu  \\ -\beta V^{\text{ext}} - \fdv{\beta \FE_\text{ex}[\rho,\alpha]}{\rho(\bm{r})} \biggr) 
\end{multline}
\begin{eqnarray}
    \label{eq:2b}
  \alpha(\bm{r}, \Omega)  &=& \frac{1}{\int \odif{\Omega}\, \psi(\bm{r}, \Omega) }\, \psi(\bm{r}, \Omega) \\
  \psi(\bm{r}, \Omega) &=& \exp \left(-\frac{1}{\rho(\bm{r})}  \fdv{\beta \FE_\text{ex,or}[\rho,\alpha]}{\alpha(\bm{r}, \Omega)} \right) \;.  \nonumber
 \end{eqnarray}

\subsubsection{Existing isotropic functionals}

{The KF potential is the sum of an isotropic hard sphere part and anisotropic patch attractions. Therefore, it appears 
promising, in a first step, to treat also the free energy functional as a sum of an isotropic hard sphere reference part and a remainder accounting for the anisotropic interactions. 
The most accurate functionals for hard spheres are based on fundamental measure theory (FMT)\cite{roth2010fundamental}. 
Here, the excess free energy becomes a function of weighted densities $n_{\nu}(\bm{r})$ --- convolutions of the 
(orientationally averaged) particle density $\rho(\bm{r})$ with a set of geometry--based kernels.
It might seem surprising, that, in a second step, the anisotropic remainder can be approximated perturbatively with a functional depending
again only on the orientationally averaged particle density $\rho(\bm{r})$.
In the bulk, where the free energy only depends on the homogeneous bulk density, such a perturbation theory was 
developed by Wertheim (theory of associating fluids) for particles interacting attractively through $N_\mathrm{p}$ bonding sites \cite{wertheim1987thermodynamic}.
Only certain classes of bonding states are permitted which helps to restrict the possible number of configurations appearing in calculations, making an analytical treatment feasible. 
For treating inhomogeneous systems, the Wertheim bulk free energy needs to be ``functionalized'', and a first suggestion based 
on FMT has been given in Ref.~\cite{yuwu2002}. 
This functional has been further refined in \cite{stopper2018bulk} (``Stopper--Wu functional'') and has shown
rather good agreement with simulations in a medium to high temperature regime.}

{The sum of the hard sphere excess free energy (here taken as the original Rosenfeld functional) and the association part of
Ref.~\cite{stopper2018bulk} define $\FE_{\mathrm{ex,iso}}[\rho]$ in the free energy splitting of eq.~\eqref{eq:22}:}
\begin{equation}
  \label{eq:23}\small
  \beta \FE_{\mathrm{ex,iso}}[\rho] = \int \odif{\bm{r}} \left[ \Phi_{\mathrm{hs}}(\left\{ n_{\nu}(\bm{r}) \right\})  +
  \Phi_{\text{bond}}(\left\{ n_{\nu}(\bm{r}) \right\}) \right]
\end{equation}
\begin{multline}
  \label{eq:29}
  \Phi_{\mathrm{hs}}(\left\{ n_{\nu}(\bm{r}) \right\}) = -n_{0} \ln(1-n_{3}) \\
  + \frac{n_{1} n_{2} - {\bm{n}_{1}\cdot \bm{n}_{2}}}{1-n_{3}} + \frac{n_{2}^{3} - 3 n_{2} \bm{n}_{2}\cdot \bm{n}_{2}}
  {24 \pi (1-n_{3})^{2}}
\end{multline}
\begin{multline}
  \label{eq:24}
  \Phi_{\mathrm{bond}}(\left\{ n_{\nu}(\bm{r}) \right\}) = N_\text{p} n_{0}(\bm{r}) \xi^{q}(\bm{r}) \\
\times  \left[ \ln X(\bm{r}) - \frac{X(\bm{r})}{2} + 1/2 \right]
\end{multline}
{For the exact definitions of the weighted densities $n_\alpha(\bm{r})$, $\xi(\bm{r})$ as well as the space-dependent
bonding probability $X(\bm{r})$, we refer to Ref.~\cite{stopper2018bulk} (Sec. IIC therein).}

\subsubsection{Anisotropic mean-field ansatz}
{A simple mean-field ansatz for the anisotropic part of the free energy functional is given by}
\begin{multline}
  \label{eq:25}
  \FE_{\text{ex}} = {\frac{1}{2}}\int \odif{\bm{r}_1}\odif{\bm{r}_2} \odif{\Omega_1} \odif{\Omega_2} \rho(\bm{r}_1)\alpha(\bm{r}_1, \Omega_1) \\ \times \rho(\bm{r}_2) \alpha(\bm{r}_2, \Omega_2) {V^{\text{MF}}}(\bm{r}_1 - \bm{r}_2, \Omega_1, \Omega_2)
\end{multline}
{This ansatz defines a yet unknown, orientation--dependent  mean-field potential $V^{\text{MF}}$ as the kernel. If one sets $V^{\text{MF}}=V^{\text{KF}}$, one obtains the standard RPA approximation for the anisotropic part in the KF fluid.
We expand both the orientations $\alpha(\bm{r}, \Omega)$ into the symmetrized basis functions $\Delta^b_{\beta' [j] }$,
\begin{equation}
   \alpha(\bm{r}, \Omega) = \sum_{b \beta' j}  \alpha^{b}_{\beta' [j]}(\bm{r})  \Delta^b_{\beta' [j]}(\Omega)
\end{equation}    
as well as the mean-field potential,
\begin{equation}
   V^{\text{MF}}(\bm{r}, \Omega_{1}, \Omega_{2}) = \sum_{m n l [ij]} M^{mnl}_{[ij]}(r) \Phi^{mnl}_{[ij]}(\Omega_{1}, \Omega_{2}, \hat{\bm{r}}),
\end{equation}
where the $\Phi^{mnl}_{[ij]}$ are given by eq.~\eqref{eq:Phi_Delta}.  With these expansions, eq.~\eqref{eq:25} becomes}
\begin{multline}
  \label{eq:26} \small
  {\FE_{\text{ex}} = \frac{1}{2} \int \odif{\bm{r}_1}\odif{\bm{r}_2}   
  \rho(\bm{r}_{1})\rho(\bm{r}_{2}) } \sum_{AB}
  \alpha^{a}_{\alpha' [i]}(r_{1}) \alpha^{b}_{\beta' [j]}(r_{2}) \\
 \times \left[ \sum_{l \lambda'} M^{abl}_{[ij]}(r_{12}) \tj{a}{b}{l}{\alpha'}{\beta'}{\lambda'} \conj{D^{l}_{\lambda' 0}}(\hat{\bm{r}}_{12}) \right] \frac{64 \pi^{4}}{c_{a}c_{b}}
\end{multline}
with the sum over $A$ standing for $\sum_{a \alpha' i}$ and for the same over $B$.

\subsubsection{Mean field with wall symmetry}
For the KF fluid between flat walls, the cylindrical symmetry in the wall plane introduces further constraints on the expansion coefficients. 
As explained above, only orientational moments with a zero in the first lower index remain, $ \alpha^{b}_{\beta' [j]} \to  \alpha^{b}_{0 [j]} $, which also helps to reduce the number of contributing $3j$ symbols. 
Densities only depend on the $z$-coordinate (perpendicular to the walls), $\rho(\bm{r}) \to \rho(z)$, and the volume element 
is split as $\odif{\bm{r}}= \odif{z} \odif{\bm{r}_{\parallel}}$. The integral over the in-plane area element $ \odif{\bm{r}_{1,\parallel}}$ gives
$A$, the area of the wall.
\begin{figure}[h]
  \centering
  \includegraphics[width=0.9\columnwidth]{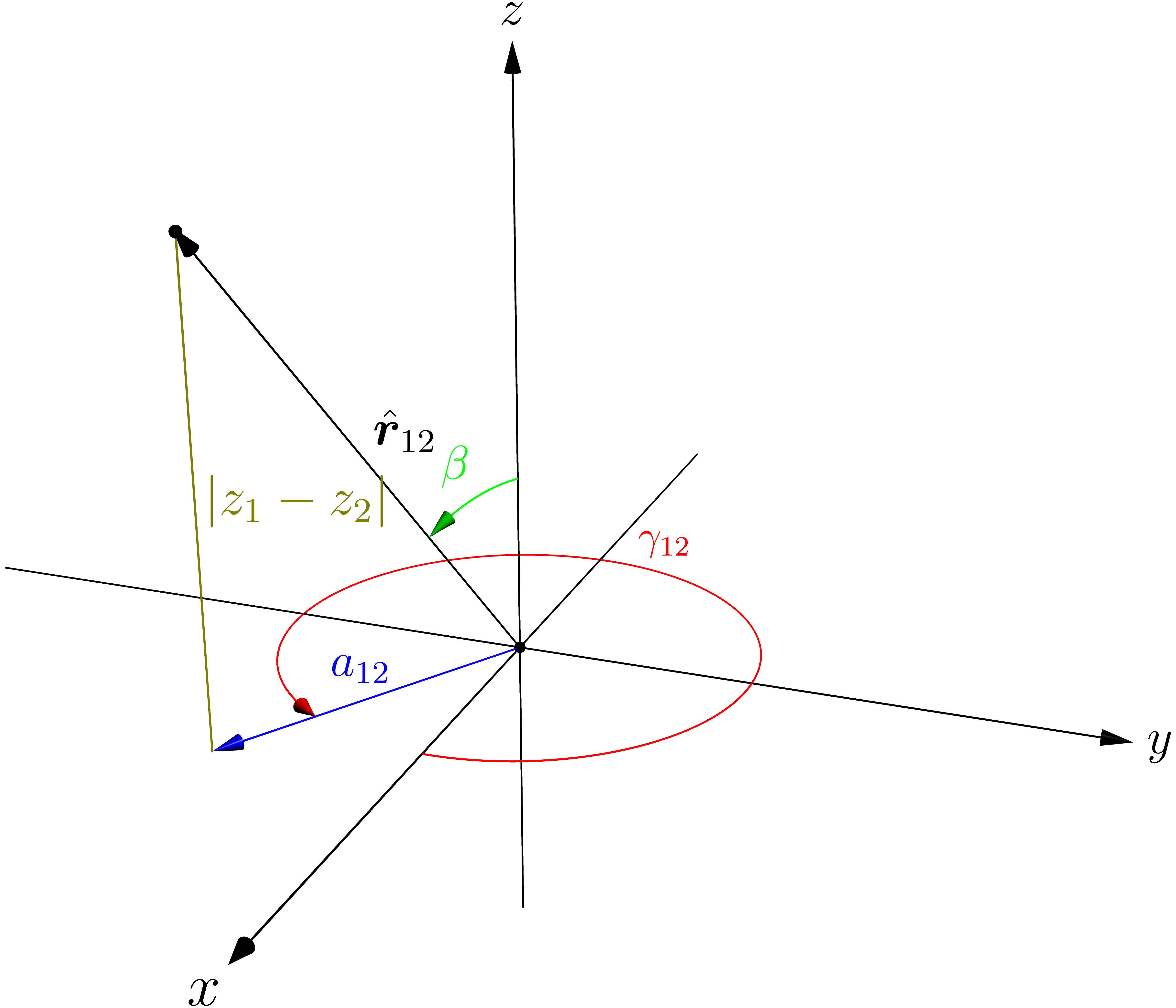}
\caption{Decomposition of the distance vector between two particles into components perpendicular and parallel to the wall. Due to the cylindrical symmetry we can integrate out $a_{12}$ and $\gamma_{12}$}
  \label{fig:cyl_integ}
\end{figure}

The second  in-plane area element $ \odif{\bm{r}_{2,\parallel}} = a_{12} \odif{a_{12}} \odif{\gamma_{12}}$ requires
integrations over the relative in-plane distance $a_{12}$ between points 1 and 2 as well as the relative azimuthal angle $\gamma_{12}$, see fig.~\ref{fig:cyl_integ}. 
We obtain

\begin{align}
    \label{eq:rpa_wall}
    \FE_{\text{ex}} =& {  \frac{A}{2} \int \odif{z_1} \odif{z_{2}} \odif{a_{12}} a_{12}  }  \odif{\gamma_{12}} \rho(z_{1})  \rho(z_{2})\nonumber\\
    &\sum_{AB}
    \alpha^{a}_{0 [i]}(z_{1}) \alpha^{b}_{0 [j]}(z_{2})\\
    &\left[ \sum_{l \lambda'} {M}^{abl}_{[ij]}(r_{12}) \tj{a}{b}{l}{0}{0}{\lambda'}
      \conj{\wD^{l}_{\lambda' 0}}(\hat{\bm{r}}_{12}) \right]  \frac{64 \pi^{4}}{c_{a}c_{b}}.\nonumber
\end{align}

The distance between the two integration points is $r_{12} = \sqrt{a_{12}^2 + |z_1 - z_2|^2}$, and by using the two angles $\beta = \arctan \frac{a_{12}}{(z_1 -z_2)}$ and $\gamma_{12}$. 
the orientation vector is given by $\hat{\bm{r}}_{12}=( \sin{\beta} \cos{\gamma_{12}}, \sin{\beta} \sin{\gamma_{12}}, \cos\beta)^\top $.
The appearing Wigner D-matrix with a lower index 0 is proportional to a spherical harmonic,
\begin{equation}
  \wD^l_{\lambda' 0}(\hbm{r}_{12}) = \sqrt{\frac{4\pi}{2 l+1}} Y^{\lambda'}_l  (\beta, \gamma_{12})\;.
\end{equation}
Thus one sees that %
the integration over $\gamma_{12}$ gives the condition $\lambda' = 0$ and the D-matrix reduces to the Legendre polynomial
\begin{equation}
 \wD^l_{00}(\hat{\bm{r}}_{12}) = P_l(\cos \beta)=  P_l \left( \frac{z_1-z_2}{\sqrt{a_{12}^2 + |z_1 - z_2|^2}} \right)  .
\end{equation}
Using this, the free energy in wall symmetry, eq.~\eqref{eq:rpa_wall}, is given by 
\begin{multline}
  \label{eq:fex_planar}
  \FE_{\text{ex}} =  \frac{A}{2}  \int \odif{z}\odif{z'} \rho(z) \rho(z') \\ \times \sum_{i'j'} \alpha^{i'}(z)   \alpha^{j'}(z') M^{i'j'}(z- z')
  \end{multline}
where the superindices $i',j'$ unite the indices of the $\alpha$-moments:
\begin{equation}
  \alpha^{a}_{0 [i]} \to \alpha^{i'}\;, \qquad  \alpha^{b}_{0 [j]} \to \alpha^{j'}\;, 
\end{equation}
and the reduced mean-field kernel $M^{i'j'}$ is given by
\begin{multline}
  M^{i'j'}(z) =  \sum_l \tj{a}{b}{l}{0}{0}{0} \\  \int \odif{a_{12}} a_{12}   M^{abl}_{[ij]}\left(\sqrt{a_{12}^2 + z^2}\right) \\
   \times P_l\left( \frac{z}{\sqrt{a_{12}^2 + z^2}}\right)\;\frac{128 \pi^{5}}{c_{a}c_{b}} \;.
 \label{eq:kernel_reduced}
\end{multline}
This result for the reduced mean-field kernel has important significance for the machine learning problem.
From simulation data in the flat wall geometry only the elements of $M^{i'j'}$  can be learned, up to a chosen cutoff
for the indices $i',j'$. However, these elements $M^{i'j'}$  only contain a subset of the moments $M^{abl}_{[\mu\nu]}$ for the  orientational invariants  of the full mean-field kernel, and moreover these in an integrated form, see eq.~\eqref{eq:kernel_reduced}. In perspective, this points to the necessity of including further types of external potentials, radially symmetric and orientation-dependent, into ML training procedures for learning a full mean-field kernel or kernels beyond mean field.

However, in the random phase approximation the reduced mean-field kernel can be computed explicitly. We expand $V^\text{KF}$ (see eq.~\eqref{eq:16}) into symmetrized orientational invariants,
\begin{equation}
 V^{\text{KF}}(\bm{r}, \Omega_{1}, \Omega_{2}) = \sum_{m n l [ij]} V^{mnl}_{[ij]}(r) \Phi^{mnl}_{[ij]}(\Omega_{1}, \Omega_{2}, \hat{\bm{r}}),
\end{equation}
and project out the kernel moments $V^{mnl}_{[ij]}(r)$, obtaining:
\begin{align}
 \label{eq:vmoments_rpa}
  V^{mnl}_{[ij]}(r) &= \epsilon^{mnl}_{[ij]} \phi_\text{sw}(r) \\
\epsilon^{mnl}_{[ij]} &=  \frac{2l+1}{256 \pi^5}\int \odif{\Omega_1} \odif{\Omega_2} \odif{ \hat{\bm{r}}_{12} } \conj{\Phi^{mnl}_{[ij]}} \\
& \times \sum_{\alpha,\beta=1}^{N_\text{p}}  \phi_{\text{p}}(\bm{r}_{12}, \hat{\bm{r}}_{1}^{\alpha}(\Omega_{1}), \hat{\bm{r}}_{2}^{\beta}(\Omega_{2})) \nonumber 
\end{align}
The $r$--dependence is entirely in $\phi_\text{sw}(r)$, 
and the dependence on all indices is contained in the numbers $ \epsilon^{mnl}_{[ij]}$ which need to be computed numerically.
Then eq.~\eqref{eq:kernel_reduced} can be applied with $M \to V$.

\subsubsection{Mean-field ansatz for machine learning}
\label{sec:ansatz}
For the actual machine learning procedure, we use the free energy in eq.~\eqref{eq:fex_planar} with the kernel $M^{i'j'}(z)$
to be determined by ML. 
The superindices $i',j'$ contain an angular momentum--like index and a second index labelling the possible moments contributing for this agular momentum.
We allow all possible interactions between symmetrized moments up to certain number $l$ for the angular momentum--like index. Apart from being a practical necessity, we also observe that with increasing $l$ the moments become smaller, further providing a rationale for this cutoff. 
We chose to include moments up to $l<6$ as extracting moments with low enough noise for higher $l$ from simulation data proved to be increasingly time-consuming. Additionally, due to symmetry, moments with $l=5$ vanish, leaving us with just two nonvanishing
superindices $=3,4$ as there is only one moment for these $l$ (see also tab.~\ref{tab:basis_fn}) . 

For the task of learning a functional for the orientational part $\FE_{\text{ex,or}}$ of the free energy
and keeping the isotropic, reference part $\FE_{\text{ex,iso}}$  fixed by the Stopper--Wu functional
(see Sec.~\ref{sec:dft_general}), we  specifically exclude interactions that only depend on two zero moments, i.e. $M^{00}$, such that the bulk behavior (where the orientational distrubution becomes isotropic) remains untouched. 
Thus the general form of the mean-field ansatz for  $\FE_{\text{ex,or}}$ for the ML problem is
\begin{multline}
\label{eq:11}
    \FE_{\text{ex,or}}^\text{mf} = \frac{A}{2} \int \odif{z}\odif{z'} \rho(z) \rho(z') \\ \times 
    \sum_{\substack{i,j \\ \text{not }i=j=0}} \alpha^i(z)   \alpha^j(z') M^{ij}(z- z')
  \end{multline}
where the indices $i$ and $j$ are the angular momentum--like indices $\{0,3,4\}$.
From eq.~\eqref{eq:kernel_reduced}, the fact that for a $3j$ symbol of the form $\tj{a}{b}{c}{0}{0}{0}$ the sum of the upper row must be even and the symmetry
properties of the Legenedre polynomials
we can deduce the symmetry properties of the reduced mean-field kernel moments for negative distances, which are shown in table~\ref{tab:sym_mf}.

\begin{table}[h]
\centering
\begin{tabular}{cc}
\toprule
  $M^{ij}$ & parity \\\midrule
  (0, 3) & odd \\
  (0, 4) & even \\
  (3, 3) & even \\
  (3, 4) & odd \\
  (4, 4) & even \\\bottomrule
\end{tabular}
\caption{Symmetry properties of the reduced mean-field kernel moments under inversions around the origin}
\label{tab:sym_mf}
\end{table}
Taken together with the fact that 
\begin{equation}
    M^{ij}(-z) = M^{ji}(z)
\end{equation}
we reduce the total number of independent mean-field moments to the five shown in table~\ref{tab:sym_mf}.
Further, the moments are assumed to be translationally invariant, i.e. they only depend on the distance between the moments
thus the integrals reduce to convolutions which can be performed efficiently using the fast Fourier transform.

The density functional derivative of this mean-field functional is
\begin{multline}
  \label{eq:fdv_rho}
 \fdv{ \FE_{\text{ex,or}}^\text{mf}[\rho,\alpha]}{\rho(z)} =  %
 \\ \sum_{i, j} \alpha^i(z)  \left[ \rho \alpha^j  \ast M^{ij} \right] + \alpha^j(z) \left[ \rho \alpha^i  \ast \overline{M}^{ij} \right]
\end{multline}
where $\overline{M}(z)$ = $M(-z)$ and the symbol $\ast$ denotes a convolution. 
For the equation that determines the orientation distribution, we need the derivative w.r.t. the individual moments
\begin{equation}
  \label{eq:fdv_alpha}\small
   \fdv{  \FE_{\text{ex,or}}^\text{mf}[\rho,\alpha]} { \alpha^k(z)} = \rho(z) \left[ \rho \alpha^j \ast M^{kj} \right] + \rho(z) \left[\rho \alpha^i \ast \overline{M}^{ik} \right]
\end{equation}

As a consequence of the mean-field ansatz, it turns out, that the self-consistent equation for the density, eq.~\eqref{eq:2}, can be further simplified. 
Upon using eq.~(\ref{eq:2b}), the entropic contribution of the orientation distribution in eq.~\eqref{eq:2}  becomes
\begin{multline}
    -\int \odif{\Omega}\, \alpha(z,\Omega) \log(K \alpha(z, \Omega)) 
     =  \\\log  \frac{1}{K} \int \odif{\Omega} \psi(z, \Omega) \\ +
    \int \odif{\Omega}\, \alpha(z, \Omega) \frac{2l+1}{K} \conj{\wD^l_{mn}}(\Omega) \frac{1}{\rho(z)} \fdv{\beta \FE^\text{mf}_\mathrm{ex,or}[\rho, \alpha]}{\alpha^l_{mn}}
\end{multline}
Plugging in our solution from \cref{eq:fdv_alpha} and comparing to \cref{eq:fdv_rho} shows that the term on the right hand side cancels the contribution of $\delta \FE_{\text{ex,or}}^\text{mf}/\delta \rho$ in the Euler-Lagrange equation \eqref{eq:2} which
becomes:
\begin{equation}
 \label{eq:rho_ML}\small
\rho(z) = \frac{\int \odif{\Omega}\, {\psi(z,\Omega)} }{K} \exp \biggl(   \beta \mu  \\ -\beta V^{\text{ext}} - \fdv{\beta \FE_\text{ex,iso}[\rho]}{\rho({z})} \biggr) 
\end{equation}
This is inherent to the mean-field ansatz and as a consequence we can not fit $\FE^\text{mf}_\text{ex,or}$ to correct the differences between our reference functional and the simulation data. The only change the orientation distribution induces on the density is through the multiplicative first term on the rhs of eq.~\eqref{eq:rho_ML}
which is already fixed by the observed orientation distribution $\alpha(z, \Omega)$ and turns out to be rather small as we will later see.
Nonetheless, this simplified approach is a valuable starting point for further more sophisticated models.

The weak influence of the mean-field $\FE^\text{mf}_\text{ex,or}$ on the orientationally averaged density profile motivates to consider separately a correction to the reference functional, $\Delta \FE^\text{mf}_\text{ex,iso}$, via the moment $M^{00}$ which has been neglected above. Only through this correction it is possible to fit better the equation of state which is reasonable for the Stopper-Wu functional (i.e. the one from Wertheim) but clearly an approximation. This explicit correction $\Delta \FE^\text{mf}_\text{ex,iso}$ brings additional issues which will be discussed in the results section, \cref{sec:orient-dens-corr}.   

\section{KF model calculations: training data and reference model results}
\label{sec:kfsim}

\begin{figure}%
  \centering
  \includegraphics[width=.9\columnwidth]{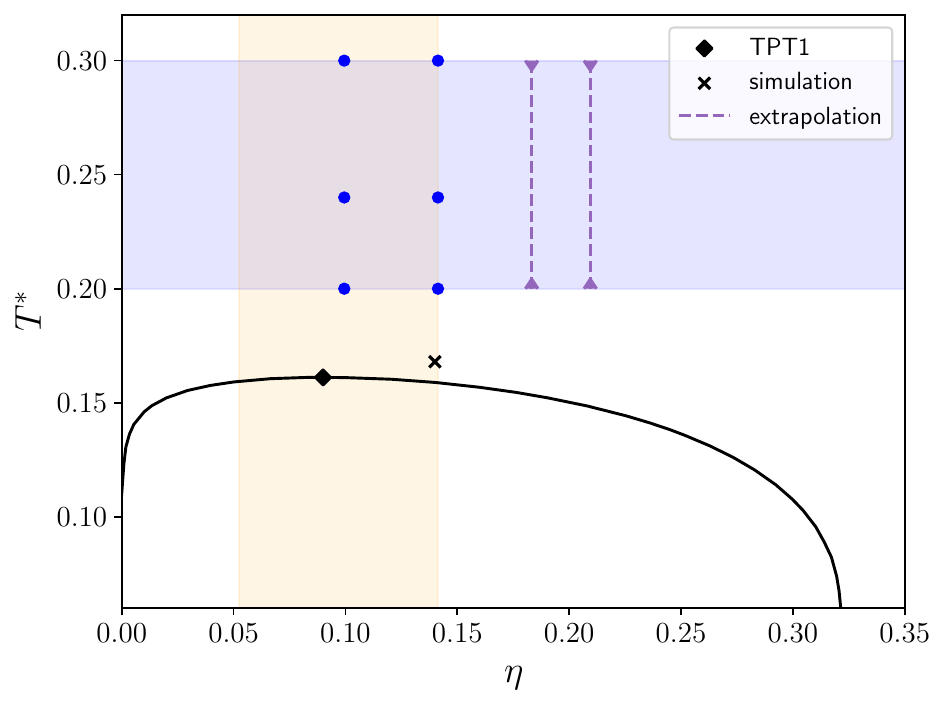}
  \caption{Phase diagram for the Wertheim theory based reference functional (TPT1). In the cross section of the shaded $T^*$ and $\eta$ intervals, training data have been generated. The blue data points correspond to the state points for the density profiles shown in \cref{fig:kf_comparison_more}. The black marker shows the location of the critical temperature and the line corresponds to the coexistence-line between the gas and liquid phase.}
  \label{fig:phasediag}%
\end{figure}

Parameters for the KF system were chosen as $N_\text{p}=4$, $\delta=0.119$ and $\cos \theta = 0.895$. For these parameters,
the critical point was determined in ref.~\cite{foffi2007possibility} with coordinates $\rho_c^*= \rho_c\sigma^3 = 0.267$
($\eta_c = (\pi/6)\rho_c^* = 0.140 $) and $T_c^* = (kT)/\epsilon = 0.1682$, using simulations. 

\subsection{Reference functional results}

\begin{figure}[t]
  \centering
    \includegraphics[width=\columnwidth]{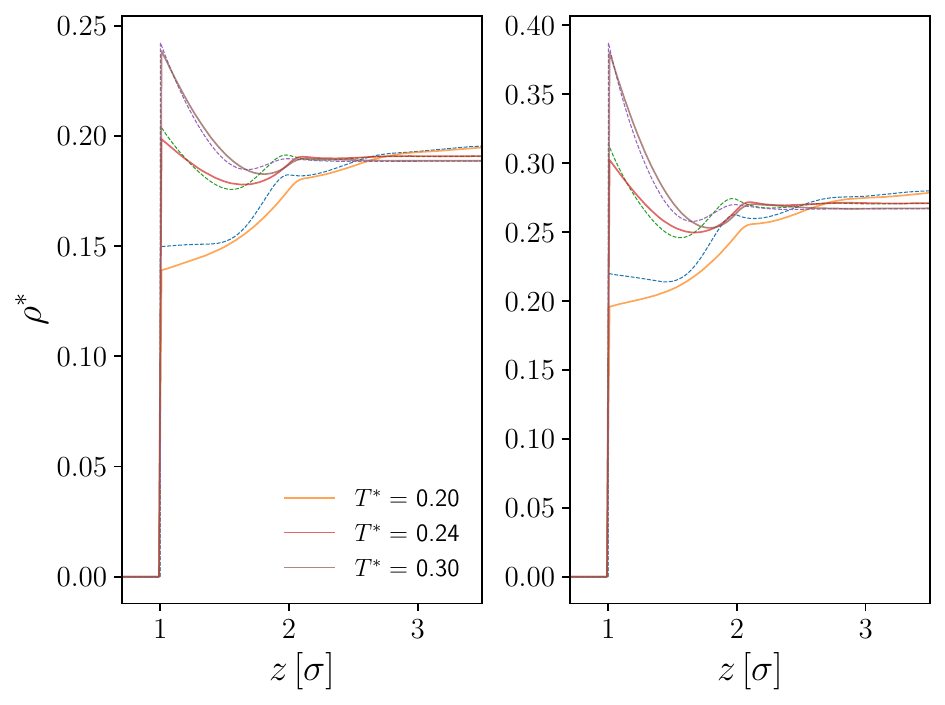}
  \caption{Comparison of data obtained through simulation of the KF model (full line) with those that result from the excess free energy in eq.~\eqref{eq:23} (dashed line), at the bulk densities $\rho_b = 0.19$  (left) and $\rho_b = 0.27$ (right) for different temperatures.}
  \label{fig:kf_comparison_more}
\end{figure}

The gas-liquid coexistence curve resulting from the reference functional \eqref{eq:23} is the one from Wertheim theory for associating liquids and can be found in 
ref.~\cite{stopper2019structure}, fig. 3.3. The critical temperatures from Wertheim theory and simulations match quite well but there is a discrepancy in $\eta_c$, see fig.~\ref{fig:phasediag}. Note, however, that the binodal is very flat in the vicinity of the critical point such that the simulation critical point is very close to the reference binodal.
In \cref{fig:kf_comparison_more}, we show profiles $\rho(z)$ of the averaged density at a hard wall obtained from the reference
functional and simulations. 
As one can see from the plots for the two densities in Fig.~\ref{fig:kf_comparison_more}, the contact density $\rho_c$ at the wall becomes lower with decreasing temperature, in accordance with approaching the binodal (onset of ``drying'') and reflected by the hard-wall theorem $\beta p= \rho_c$ (where $p$ is the pressure of the bulk fluid). Both simulation and reference functional results show this behavior, however, with lower temperature
the agreement between theory and simulation worsens. It is clear, that preferred particle orientations start to become more and more important in shaping the fluids behavior.

\subsection{Simulations}
In order to obtain the necessary training data of the orientation distributions, we
simulated the system as a canonical Monte Carlo simulation (constant $NVT$ ensemble), using a code from Rovigatti et al., described in ref.~\cite{rovigatti2018simulate}. The code was suitably modified to incorporate hard walls at distance $L_\text{w}$ with
their normal in $z$--direction. Initial configurations are generated by randomly 
placing particles into the simulation box. The number of particles was
fixed to $N=1000$ and the length of the cubic simulation box is hence
$L=\left( N/\rho \right)^{1/3}$. After the initial thermalization phase we 
save a sample of the system every 400 MC sweeps.
For training state points we chose temperatures $ T^* \in \{0.20;0.22;0.24;0.26;0.30\} $  and bulk densities $\rho^* \in \{0.10;0.13;0.16;0.19;0.21;0.24;0.27\} $. See also fig.~\ref{fig:phasediag} for the location of the training points in the
phase diagram.

From equilibrated samples, profiles $\rho(z)$ of the averaged density are obtained by standard histograms and orientation profiles are obtained according to eqs.~\eqref{eq:expansion_coeff} and \eqref{eq:product_ansatz} as follows
\begin{equation}
  \label{eq:alpha_hist}
  \alpha^l_{m[j]}(z_i) = \frac{1}{\rho(z_i)} \frac{2l + 1}{K} \sum_{\Omega_k \in z_i} \Delta^l_{m[j]}(\Omega_k)
\end{equation}
where $z_i$ labels the bin the particle is in and $\Omega_j$ is the orientation of the $j$th particle in bin $z_i$.

\begin{figure}%
  \centering
  \includegraphics[width=\columnwidth]{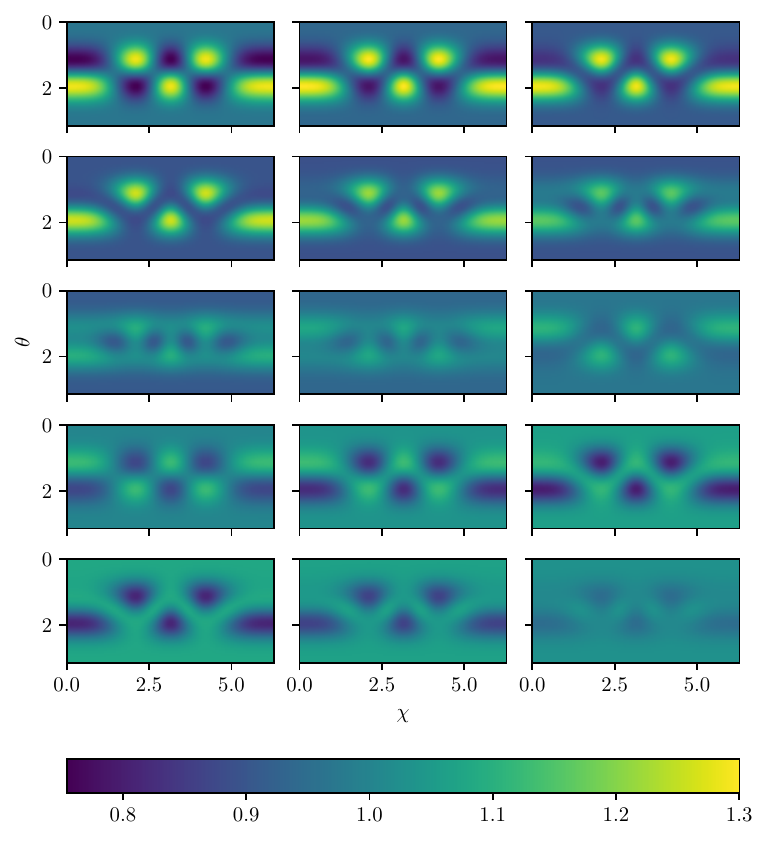}
  \caption{Each image shows the probability distribution $\alpha(z, \theta, \chi)$ at a certain $z$, where the angle $\phi$ drops out due to the cylindrical symmetry. For easier reading, $\alpha$ is scaled by $K$ such that
  in the bulk it reduces to one. The first picture starts right at the wall $z=\sigma$ and each subsequent one (left to right and row-wise top to bottom) is displaced by $0.07\sigma$ in the $z$ direction. The state point is at $\rho^*_b = 0.27$ and $T^*=0.2$, displaying the strongest anisotropy in the dataset.}
  \label{fig:alpha_odf}%
\end{figure}

\begin{figure}[ht!]
\begin{tikzpicture}[scale=1.3]
\begin{scope}[3d view={128}{12}]
    \filldraw[draw=red!30, fill=red!10] (1, 1, 0) -- (-1, 1, 0) -- (-1, -1, 0) -- (1, -1, 0) -- cycle;

    \draw[->] (-1, 0, 0) -- (1, 0, 0) node[pos=1.1] {$x$};
    \draw[->] (0, -1, 0) -- (0, 1, 0) node[pos=1.1] {$y$};
    \draw[->] (0, 0, 0) -- (0, 0, 2) node[pos=1.1] {$z$};
        
    \begin{scope}[shift={(0, 0, 1.4)}]
    \node (A) at (0, 0, -1) [circle, draw=blue, fill=blue!50] {};
    \draw (0, 0, 0) -- (A);
    
     \node (B) at (-0.94, 0, 0.34) [circle, draw=blue, fill=blue!50] {};
    \draw (0, 0, 0) -- (B);
    
     \node (C) at (0.474, -0.82, 0.33) [circle, draw=blue, fill=blue!50] {};
    \draw (0, 0, 0) -- (C);
    
     \node (D) at (0.474, 0.82, 0.33) [circle, draw=blue, fill=blue!50] {};
    \draw (0, 0, 0) -- (D);
     \end{scope}
\end{scope}
\end{tikzpicture}
\hspace{.2cm}
\begin{tikzpicture}[scale=1.3]
\begin{scope}[3d view={128}{12}]
    \filldraw[draw=red!30, fill=red!10] (1, 1, 0) -- (-1, 1, 0) -- (-1, -1, 0) -- (1, -1, 0) -- cycle;

    \draw[->] (-1, 0, 0) -- (1, 0, 0) node[pos=1.1] {$x$};
    \draw[->] (0, -1, 0) -- (0, 1, 0) node[pos=1.1] {$y$};
    \draw[->] (0, 0, 0) -- (0, 0, 2) node[pos=1.1] {$z$};
        
    \begin{scope}[shift={(0, 0, .7)}]
    \node (A) at (-.944, 0, -.33) [circle, draw=blue, fill=blue!50] {};
    \draw (0, 0, 0) -- (A);
    
     \node (B) at (0, 0, 1) [circle, draw=blue, fill=blue!50] {};
    \draw (0, 0, 0) -- (B);
    
     \node (C) at (0.467, -0.82, -0.33) [circle, draw=blue, fill=blue!50] {};
    \draw (0, 0, 0) -- (C);
    
     \node (D) at (0.467, 0.82, -0.33) [circle, draw=blue, fill=blue!50] {};
    \draw (0, 0, 0) -- (D);
     \end{scope}
\end{scope}
\end{tikzpicture}   
\caption{The two configurations with the highest (left) and lowest (right) probability close to the wall. The one the l.h.s.\ is more likely due to the number of patches pointing towards the fluid and therefore possible bonding partners. The wall is indicated by the red square in the $xy$ plane.}
\label{fig:maxp_config}
\end{figure}
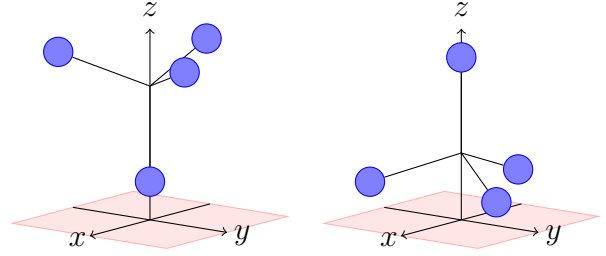

\begin{figure}[h]
  \centering
  \includegraphics[width=\columnwidth]{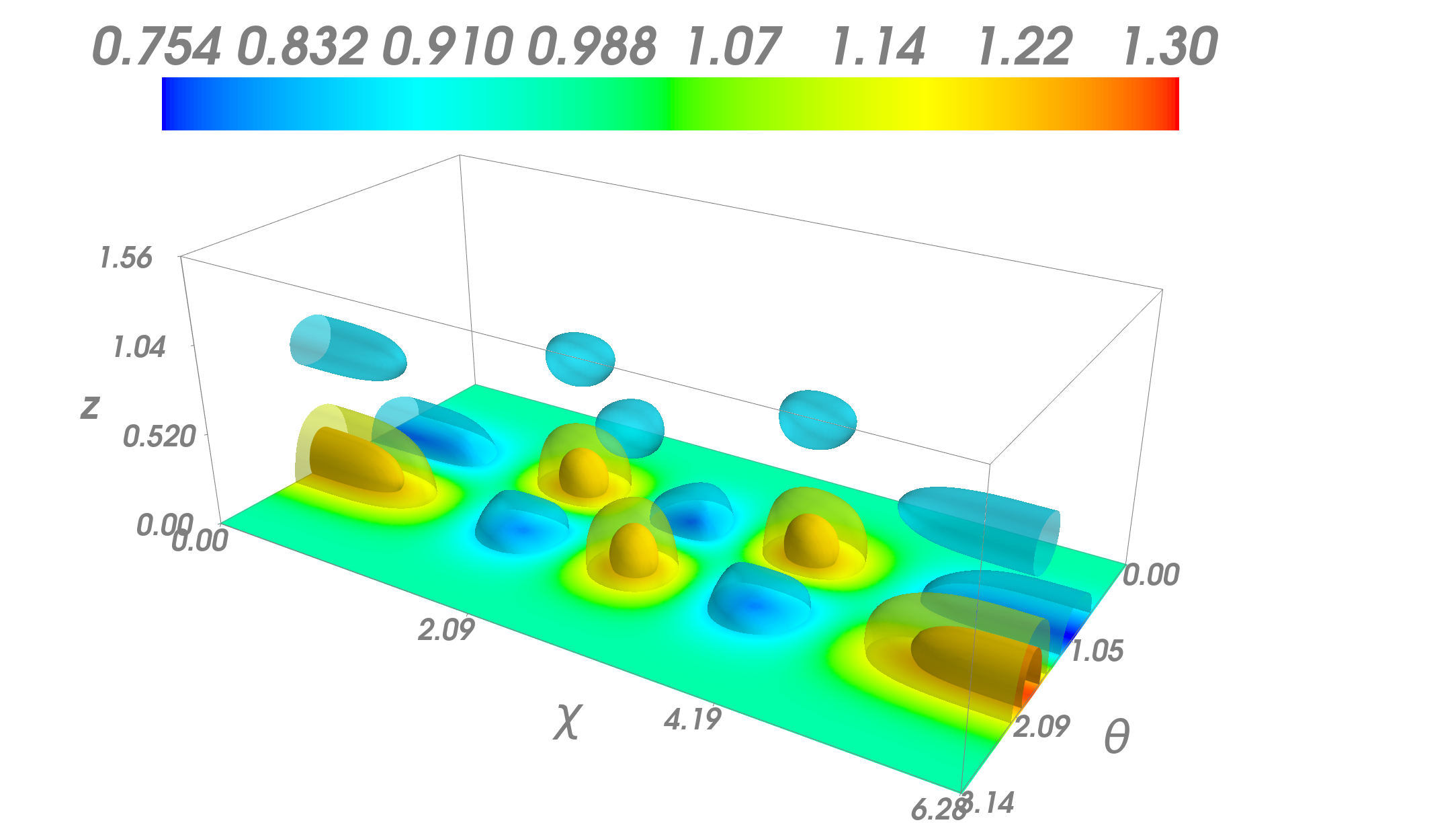}
  \caption{Three dimensional plot of the angular distribution function $\alpha(z, \theta, \chi)$ close to the hard wall for $T^*=0.2$ and bulk density $\rho^*=0.27$. The $z$ axis shows the distance from the wall. }
  \label{fig:odf3d}
\end{figure}

\begin{figure}%
  \centering
  \includegraphics[width=\columnwidth]{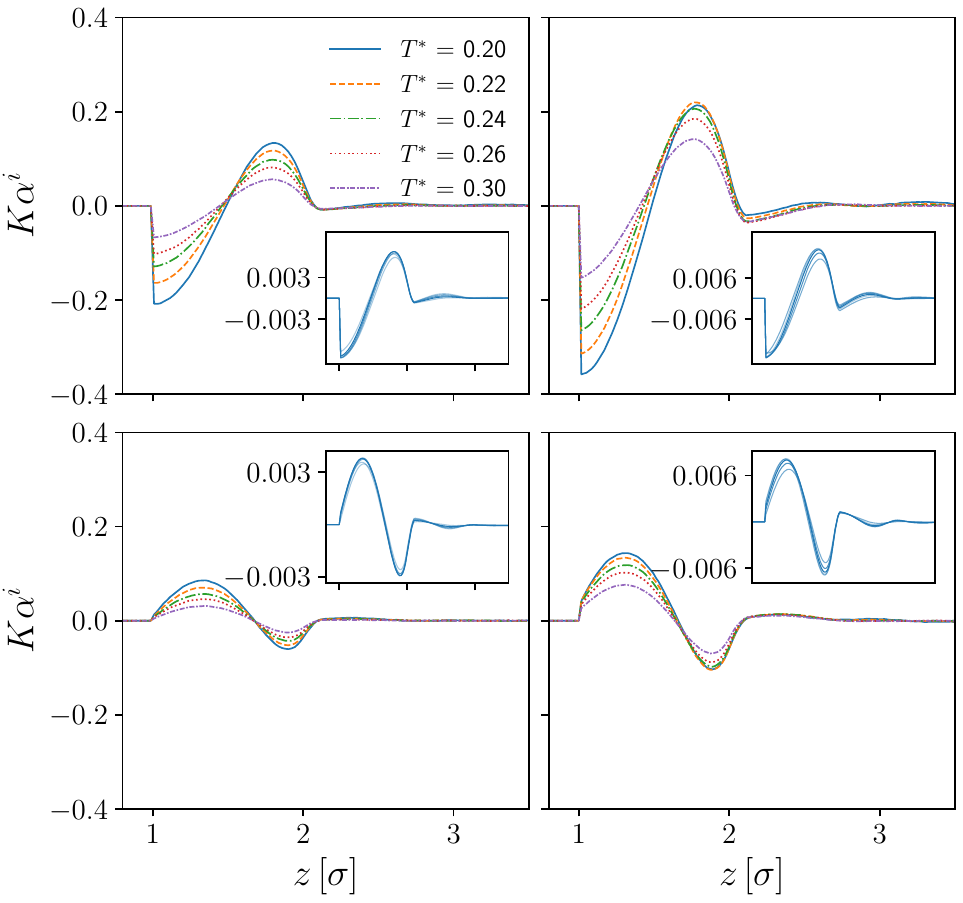}
  \caption{Here we show both orientational moments separately for different temperatures and the bulk densities $\rho_b = 0.13$ (left) and $\rho_b = 0.27$ (right). 
  The upper panels show $\alpha^3$ and the lower ones $\alpha^4$. The inlays show the orientations that result in the random phase approximation, which are off by an order of magnitude.}
  \label{fig:ori_data_same_density}%
\end{figure}

In the limit of $T^*\to\infty$ the KF model reduces to the hard sphere model and there is no anisotropy in the density profiles.
With decreasing temperature however, the bonding mechanism becomes more important and changes the orientational distribution of the fluid near the wall.
A comparison of the leading orientational moments $\alpha^3(z)$ and $\alpha^4(z)$  for different temperatures is shown in \cref{fig:ori_data_same_density}. The insets show the corresponding orientational moments as obtained from minimizing
the orientational RPA functional (mean-field kernel \cref{eq:kernel_reduced} obtained directly from the KF potential moments \cref{eq:vmoments_rpa}) at the hard wall. The RPA orientational moments are similar in shape but consistently more than an order of magnitude smaller than the moments from simulations. This points to strong orientational order at the hard wall.
Further support for the enhanced strength of orientational order comes from the explicit orientational distribution function as a function of the two nontrivial angles $\theta$ and $\chi$  (shown in  \cref{fig:alpha_odf} and also \cref{fig:odf3d} as a three dimensional plot) which can readily reconstructed from the moments. (Normalization is chosen such that it is 1 in the bulk fluid.)
We infer that the fluid features a rather strong anisotropy close to the wall, with the most probable angles being around $70\%$ more likely than the least probable ones.
As one would expect from tetrahedral symmetry, the probability distribution's maxima and minima appear four times each, once for every equivalent configuration of the patches. From \cref{fig:alpha_odf} or \cref{fig:odf3d} one notices that close to the wall there are both excluded and preferred orientations, but further away mostly exclusion effects play a role. The configurations with the highest and lowest probability are centered around two orientations (see \cref{fig:maxp_config}) between which the fluid switches back and forth, depending on the distance $z$.

\section{Machine learned DFT}
\label{sec:ml}

The goal is to find a reduced mean-field kernel $M^{ij}$ which guarantees stable numerical solutions of the EL equations, \cref{eq:2} and~\cref{eq:2b},  yielding orientational distributions and corrections to the average density profiles close to those of the Monte Carlo simulations of the system. The ``fitting'' procedure is quite complex owing to the large number of unknowns and the implicit nature of the EL equations, necessitating the use of Machine Learning methods.

In previous work, ML had been used 
to find a (symbolic) expression for the excess free energy of a 1D hard-rod and a Lennard-Jones system \cite{shang2019classical}.
There, the self-consistency of eq.~\eqref{eq:2} was exploited to find the ideal parameters $\theta$ of the network. Specifically,  
\begin{equation}
   \label{eq:ml_selfconsistency}
  \rho^*(\bm{r}; \theta)_\text{ML} = \exp \left( \beta \mu - \beta V^\text{ext}  - \beta \fdv{\FE^\mathrm{ex}[\rho; \theta]}{\rho(\bm{r})}_{\rho_\mathrm{eq}}\right)
\end{equation}
with the loss function
\begin{equation}
  L(\theta) = \int \odif{\bm{r}} \left( \rho^*(\bm{r}; \theta)_\text{ML} - \rho^*_\text{eq}(\bm{r}) \right)^2
\end{equation}
which is minimized.
Here, $\rho_\text{eq}$ is the ``ground truth'' equilibrium density profile generally obtained from simulations. %
We rewrite the above fixpoint condition as
\begin{equation}
  \label{eq:phi_root_rho}
    \Phi(\rho; \theta) = f(\rho_\text{eq}; \theta) - \rho
\end{equation}
with $f$ depending also on $\beta, \mu$ and $V^\text{ext}$,
such that the learned solutions (which depends on the parameters $\theta$ of the network) becomes the root of $\Phi$ (under the assumption that such a root exists and and is physical)
\begin{equation}
   \label{eq:rootrho}
    \rho(\bm{r}; \theta)_\text{ML} = \Root \Phi(\rho; \theta)
\end{equation}
For the present problem, the orientational distribution $\alpha$ enters as an additional argument for the function $f$ above:
\begin{equation}
  \label{eq:phi_root}
    \Phi(\rho; \theta) = f(\rho_\text{eq}, \alpha_\text{eq}; \theta) - \rho
\end{equation}
The solution of eq.~\eqref{eq:2b} for the orientational distribution  has the same structure: 
\begin{eqnarray}
    \Psi(\alpha; \theta) &=& g(\rho_\text{eq}, \alpha_\text{eq}; \theta) - \alpha \\
     \alpha(\bm{r}; \theta)_\text{ML} &=& \Root \Psi(\alpha; \theta) \label{eq:rootalpha}
\end{eqnarray}
Note, that the inputs to the functions $f$ and $g$ are the ground truth, equilibrium values.
If the ground truth is exact (i.e, no noise in the data) and the above roots fulfill the condition
$\rho_\text{ML}=\rho_\text{eq}$ and $\alpha_\text{ML}=\alpha_\text{eq}$, then these are also simultaneous roots of the 
following modified functions:
\begin{gather}
\Phi(\rho, \alpha; \theta) = f(\rho, \alpha; \theta) - \rho  \nonumber \\
\Psi(\rho, \alpha; \theta) = g(\rho, \alpha; \theta) - \alpha \label{eq:newPhiPsi}
\end{gather}
These are nothing but the original EL equations, note that the equilibrium ground truth does not appear on the rhs.  In real life however, the ground truth from simulations is noisy and the set of ML parameters $\theta$ is insufficient such that after loss minimization (using eqs.~\eqref{eq:rootrho},\eqref{eq:rootalpha})  the roots of \eqref{eq:newPhiPsi} (the solution of the EL equation for the ML functional) may deviate strongly from the ground truth or do not exist at all.
Specifically for noisy ground truth equilibrium data, when training the network using eqs.~\eqref{eq:rootrho},\eqref{eq:phi_root}, we need to solve for the root up to a rather high relative error
as we only have information of the function at the physical fixed point itself. This becomes problematic and we are eventually over-fitting, leading to an unphysical solution.

One way to alleviate these problems is to complete the whole fixed point finding procedure (finding the roots of \eqref{eq:newPhiPsi} iteratively) also during the training step, starting from initial values that are far from equilibrium values. The resulting solutions would
then be automatically self-consistent.
Although this is possible in principle, there are some drawbacks, mainly due to
the fact that we need to store the complete computational graph of the iterations with intermediate values in order to be able to back-progate through the loss and find
better parameters $\theta$. The memory consumption grows with the number of iteration steps that we do to find the root and hence we are going to be
limited to some maximum number of steps that we can do.
The further we are from the solution, the more steps are usually necessary, so one could keep memory consumption low by
starting the root-finding procedure from a point not too far from the equilibrium.
Another approach that is gaining popularity in the machine learning community is using implicit differentiation during the backward pass.
It allows one to compute the gradient w.r.t.\ the loss without retaining the complete computational graph necessary to find the fixpoint. The knowledge of
the fixpoint is enough to compute the gradient $\nabla_\theta(\cdot)$, so one can use \emph{any} method to compute the fixpoint without the need of storing intermediate value~\cite{chang2022object, bai2019deep}.

In our case, this approach translates to the following. For simplicity, we describe the method for the first investigated case of a fixed reference system with density $\rho_\text{ref}$.  We initialize the parameters $\theta_0$, e.g.\ by setting them to zero. We then find the root
of $\Psi(\alpha; \theta_0)$, starting from the intial values %
$\alpha_\text{init}$. Under the assumption that the iteration converges, we arrive at the fixpoint $\alpha_0(z, \Omega)$, where the index reminds us of the fact that this solution depends on $\theta_0$. Unlike the previous strategy, the FP might very well
be something completely different from our sought solution $\alpha_\text{eq}$. However, what we gain is the certainty that $\alpha_0$ is indeed a self-consistent solution during the whole iterative process. Making use of implicit differentiation, we can compute $\nabla_\theta \ell$, where
\begin{equation}
  \label{eq:fp_loss}
  \ell = \MSE \left( \alpha_0, \alpha_\text{eq} \right)
\end{equation}
and use this value in our optimizer to arrive at a better value $\theta_1$ and so on (MSE is the mean squared error summed over all values).

One drawback of this approach is that during the backward pass it is necessary to compute the matrix inverse of the Jacobian of the forward transformation. This is due to how automatic differentiation works internally.
Assume that $\alpha^\star$ is a fixed point of $g(\rho_\text{ref}, \alpha; \theta)$, and we are interested in how the fixed point changes upon changing the parameters of the network. Using the implicit function theorem we have
\begin{equation}
  \label{eq:31}
  \pdv{\alpha^\star}{\theta} = \pdv{g}{\theta}_{\alpha^\star} \left[ \mathbb{I} - \pdv{g}{\alpha}_{\alpha^\star} \right]^{-1}
\end{equation}
The inverse on the r.h.s. is usually solved iteratively but depending on the function $g$ and the fixed point which we are momentarily at, the solution might become increasingly expensive,
numerically unstable or even non-existent. One proposed method to alleviate this problem is the so-called \emph{Phantom Gradient}. The idea is to expand the Jacobian inverse in a Neumann series and truncate after the first term. This corresponds to replacing the inverse with the identity matrix~\cite{geng2021training}. Translated into our training procedure this means the following. We still find the fixed point
solution $\alpha^\star$ for a certain set of parameters $\theta$, however we do not take the computations into account when constructing the computational graph\footnote{Using for example \texttt{Tensor.detach()} or \texttt{jax.lax.stop\_gradient(x)} in the two popular ML packages PyTorch and JAX}. Since we know that this point is a fixed point up to a certain tolerance, we can plug it back into the function $g$ and use this computation for the backward pass. In addition to being much faster, it has been shown to improve the stability of the training \cite{chang2022object,geng2021training}.

 Note that self-consistency of the ML functional during the training was not needed in the 1D cases of refs.~\cite{shang2019classical,shang2020FEQL} and the 3D hard sphere case of ref.~\cite{sammuller2023neural}.
 There, the final ML functionals gave stable and accurate solutions of the EL equation, presumably also due to the availability of low noise simulation data in these simple systems. In ref.~\cite{cats2021} 
 (treating 3D Lennard-Jones in a generalized mean-field fashion), self-consistency of the ML functional during the training was fully maintained and error propagation was handled by explicitly calculating the derivatives w.r.t.\ $\theta$ arising from their generalized mean-field ansatz, using the properties of the solution of the minimizing EL equation. Here, our approach appears to be generalizable to more complex functional parametrizations. The problem of self-consistency of the ML functional for electron quantum DFT was also investigated in ref.~\cite{lili2021} where the minimizing EL equations are known as the Kohn-Sham (KS) equations. It was found that maintaining self-consistency (even with just a limited number of iterations during the learning procedure) acts as an additional regularizer for the ML fitting, it was called by the authors the ``KS regularizer''.

To complete the formulation of the ML procedure in our case, we describe the iterative step (i.e.\ application of the function $g$) in more detail.
It is performed as follows.
Training data consists of a set $\{\rho(z), \alpha^{3}(z) \equiv \alpha^{3}_{0[1]}, \alpha^4(z) = \alpha^{4}_{0[1]}\}$ for every bulk density and temperature.
The reduced kernels $M^{ij}(z)$ in eq.~\eqref{eq:fdv_rho} and \eqref{eq:fdv_alpha} are parametrized by $N_k$ variables each, which correspond to their value in real space on a grid from $-L/2$ to $L/2$ with spacing $\Delta z$ and they from the set $\{\theta \}$ of the ML parameters. As the spacing $\Delta z$ for the potentials is different from the spacing of the training data points we need to translate from one support to the other. This is
done by linear interpolation of the kernel variables to every $z$ axis in the data set. 
There is some freedom in the choice of $L$ and $N_k$. For one, the spacing $\Delta z$ should not be smaller than the smallest spacing present in the training data set, since we would not have accurate information at this scale. Further, we know that
the range of the mean-field potentials that originate from the inter-particle potential is finite and most likely (judging by the angle distributions) short ranged. Choosing a value of $L$ that is unnecessarily large makes the training more expensive and might induce unwanted long-range effects that are not present in the physical system.

The actual iterative step consists of\\[1em]
1. Evaluating the non-normalized orientation distribution by
  \begin{align*}
    &\psi(z,\Omega;\theta) =\\&\  \exp \left(- \frac{1}{\rho(z)} \sum_{lmn}\fdv{\beta \FE_{\mathrm{ex}}[\rho,\alpha;\theta]}{\alpha^{l}_{m[j]}(z)} \conj{\Delta^{l}_{m[j]}}(\Omega) \frac{2l + 1}{K} \right)
  \end{align*}
2. Normalizing 
  \begin{equation*}
  \alpha(z,\Omega;\theta) = \psi(z,\Omega;\theta)/\int\odif{\Omega} \psi(z,\Omega;\theta)      
  \end{equation*}
3. Projecting out the moments of interest
  \begin{equation*}    
  \small
\!\!\!\alpha^{l}_{m[j]}(z;\theta) = \frac{2l+1}{K} \int \odif{\Omega} \alpha(z,\Omega;\theta) \conj{\Delta^{l}_{m[j]}}(\Omega)
  \end{equation*}

\paragraph*{Remarks on hyperparameters for learning the kernels}
When parametrizing the reduced mean-field kernels (which are convolution kernels) directly in real space, they tend to become rather noisy during training due to the stochastic nature of the optimization
and the fact that we interpolate the potentials onto the support of the training data sets. To reduce noise, we ``coarse grain'' the kernels by defining them on a grid with larger spacing $\Delta z$
compared to the spacings of the training data.
Further, in order to regularize the kernels, we also add a $L^1$ penalty (absolute value norm) term to the loss
\begin{equation}
  \label{eq:tot_loss}
  \ell = \MSE \left( \alpha^\star, \alpha_\text{eq} \right) + \lambda \sum_{ij} L^1(M^{ij}).
\end{equation}
The stronger the parameter $\lambda$, the more expensive it becomes for the network use non-zero kernel values and hence create interactions between the moments. If the regularizer is too large we expect
to get a bad reproduction of the orientational moments, whereas for values of $\lambda$ too small, the kernels are overfitting on the noise that is
present in the training data.
To find the ideal values of $N_k$ and $L$ we did a hyperparameter-optimization with $N_k \in [64, 128]$, $\lambda$ and $L \in [4.0, 6.0, 8.0]$. We observe that the final loss does not vary
strongly with the parameters $L$ and $N_k$. The resulting kernels are smoother for the small $N_k$ but some resolution is lost in interesting regions like inside the sphere $r < \sigma/2$.

In the second case of interest, the mean-field correction to the reference functional with kernel $M^{00}$ was considered. Here, 
the roots of the coupled set of equations \eqref{eq:newPhiPsi} need to be determined in each training step. The method outlined in the previous paragraphs, however, remains unchanged, only $\alpha$ needs to be interpreted as the combined set $\{\rho,\alpha\}$.

\section{Results from ML}
\label{sec:resultsML}

\subsection{Orientational correlations for the fixed reference system}
\label{sec:sub_fixed}
\begin{figure}[t]
  \centering
  \includegraphics[width=\columnwidth]{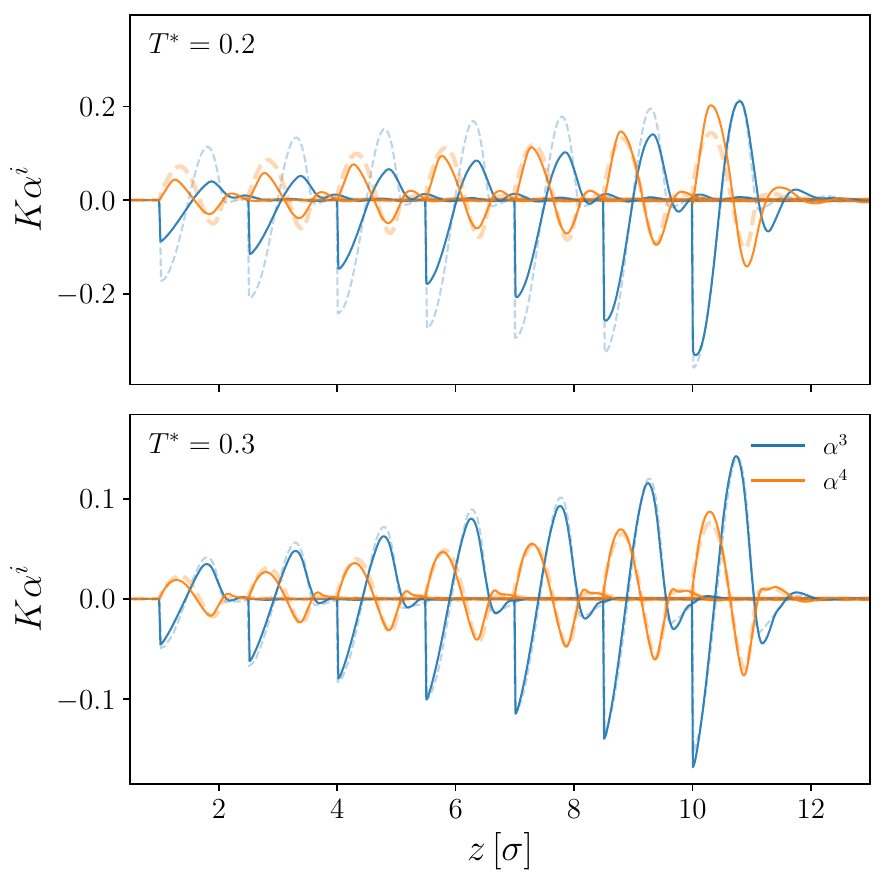}
  \caption{Leading orientational moment profiles from ML-$\FE_{\text{ex,or}}$, together with the fixed Stopper-Wu reference system.
 Bulk densities are from the set $\{0.1,\, 0.13,\, 0.16,\, 0.19,\, 0.21,\, 0.24,\, 0.27\}$, labeled with $i=0...6$.
 Profiles for different bulk densities $\rho_i$ are plotted with an offset $i\cdot 1.5\sigma$ in the $z$ direction.}
  \label{fig:ori_fixed_ref} 
\end{figure}

As described in \cref{sec:ansatz} we first have trained ML-$\FE_{\text{ex,or}}$ on each isotherm separately with no correction for the 
reference part $\FE_{\text{ex,iso}}$, i.e. setting the reduced kernel component $M^{00} = 0$. 
The resulting self-consistent orientations for different densities are shown in \cref{fig:ori_fixed_ref}, for both the highest ($T^*=0.3$) and
lowest temperature ($T^*=0.2$) in the training set. As one would expect, the agreement between ML-DFT and simulation is better for the
higher temperature, where the mean-field approximation should become more accurate (and is indeed seen to be reliable with the ML-trained kernel). %
One point that needs to be stressed is that we use the fixed reference functional to produce the density profiles, which are
known to be different from those in the simulation. As the density distribution is implicitly part of the orientation distribution
we do expect deviations w.r.t. the simulation results. These limitations notwithstanding, we see that procedure was able to correct for
this inconsistency reasonably well. For the lower temperature $T^*=0.2$ (closer to the critical temperature) there are systematic problems
for lower densities where ML-DFT produces orientational correlations smaller than the simulated ones. For higher densities these
deviations become smaller. 

\begin{figure}[h]
  \centering
  \includegraphics[width=\columnwidth]{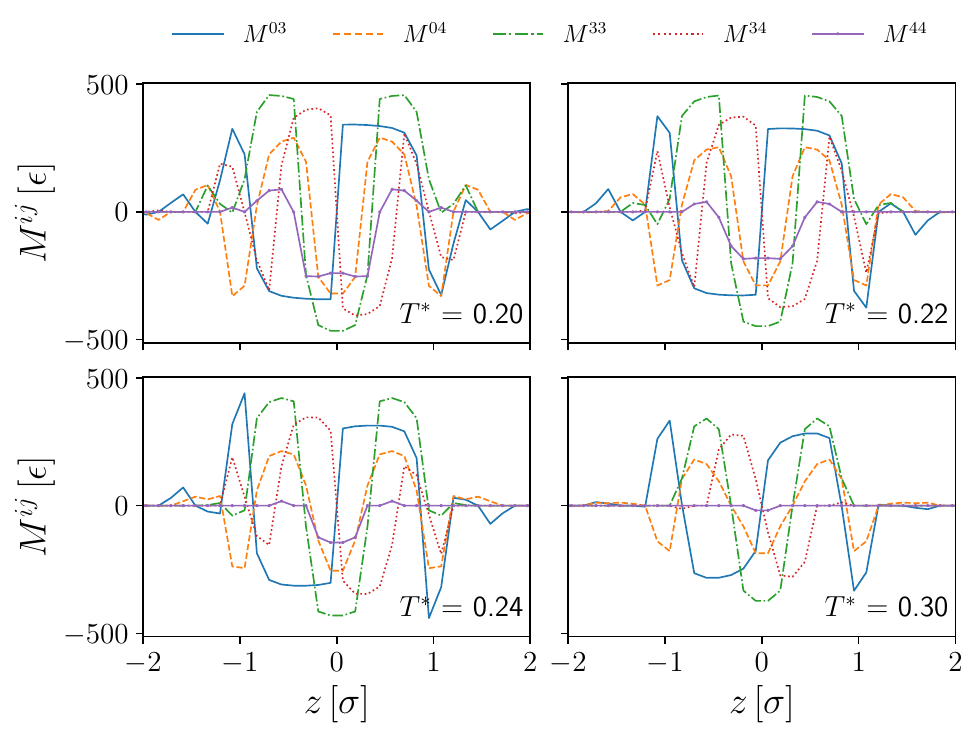}
  \caption{Learned reduced mean-field kernel moments $M^{ij}$  (in units of $\epsilon$) at different isotherms. The $L^1$ regularization strength is the same for all cases, here  $\lambda = $ \num{5e-5}}
  \label{fig:fixed_ref_isotherm}
\end{figure}

\begin{figure}[h]
  \centering
  \includegraphics[width=.9\columnwidth]{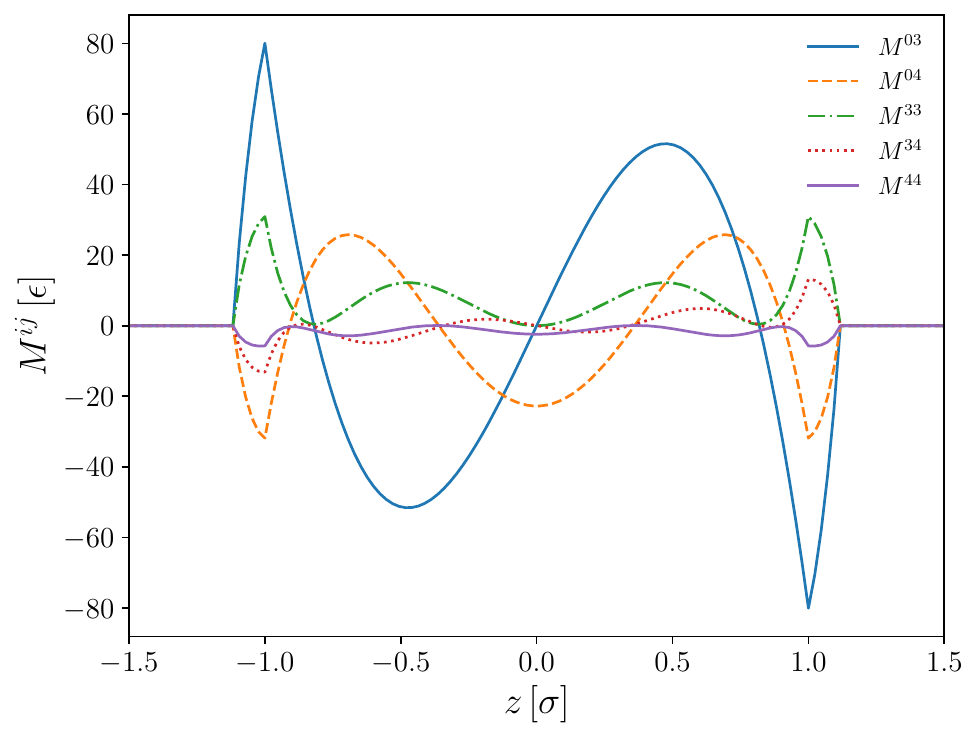}
  \caption{The  reduced mean-field kernel moments $M^{ij}$ (in units of $\epsilon$) in the RPA approximations, calculated according to \cref{eq:kernel_reduced} and using Monte Carlo integration to compute
  the moments of the Kern-Frenkel potential}
  \label{fig:rpa_ori}
\end{figure}

The associated reduced mean-field kernel moments $M^{ij}$ from the ML optimization are shown in \cref{fig:fixed_ref_isotherm}.
It is useful to compare them to the corresponding RPA moments shown in \cref{fig:rpa_ori}. Since the patch-patch attraction is square well-like, the orientational moments of the KF potential, eq.~\eqref{eq:vmoments_rpa}, are just numbers and the $z$-dependence entirely comes from the integration in eq.~\eqref{eq:kernel_reduced} with the Legendre polynomial integrand.
We see that, qualitatively, the strongest moments, $M^{03}$ and $M^{04}$, are similar in shape (but smaller in RPA), while the others show a rather different structure and the magnitude of the ML moments is much bigger. 
Furthermore, the ML moments are temperature-dependent (in contrast to the RPA moments) but decrease only slowly in magnitude with increasing temperature such that accuracy of the orientational RPA can only be expected for high temperatures.

\subsection{Orientational and density correlations for a reference system with mean-field correction}
\label{sec:orient-dens-corr}

\begin{figure}[t]
  \centering
  \includegraphics[width=\columnwidth]{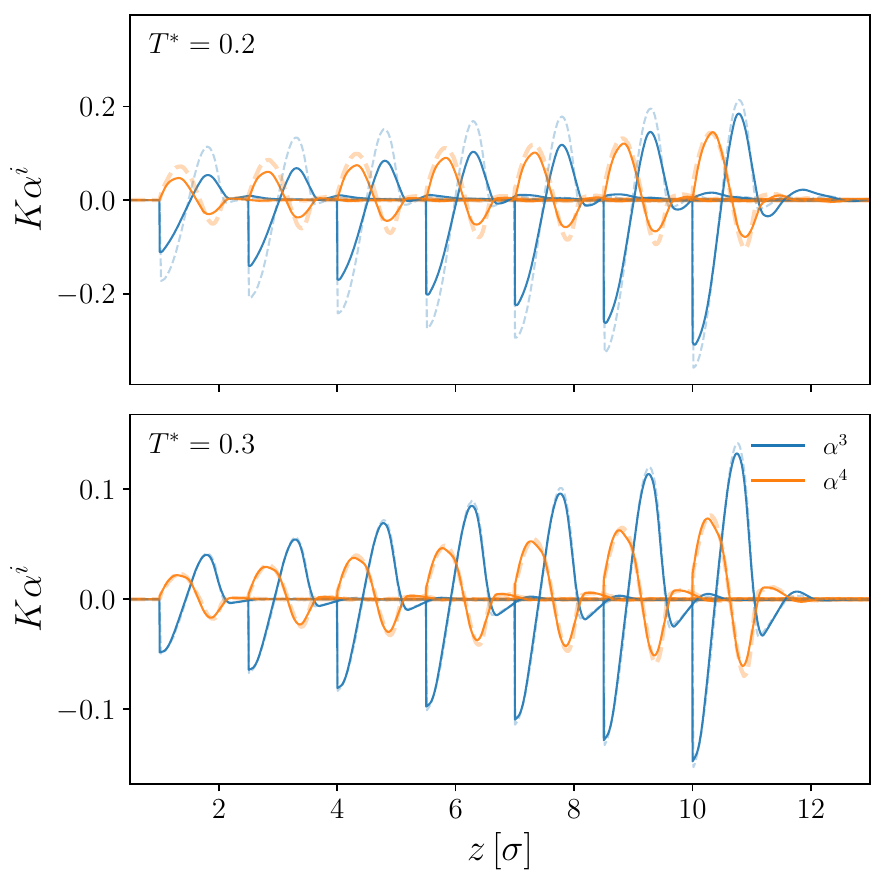}
  \caption{Resulting orientational moments from  (training  both on $\rho$ and $\alpha$). Offsets for different bulk densities as in fig.~\ref{fig:ori_fixed_ref}. 
  }
  \label{fig:ori_mf_correction}
\end{figure}

\begin{figure}[t]
  \centering
  \includegraphics[width=\columnwidth]{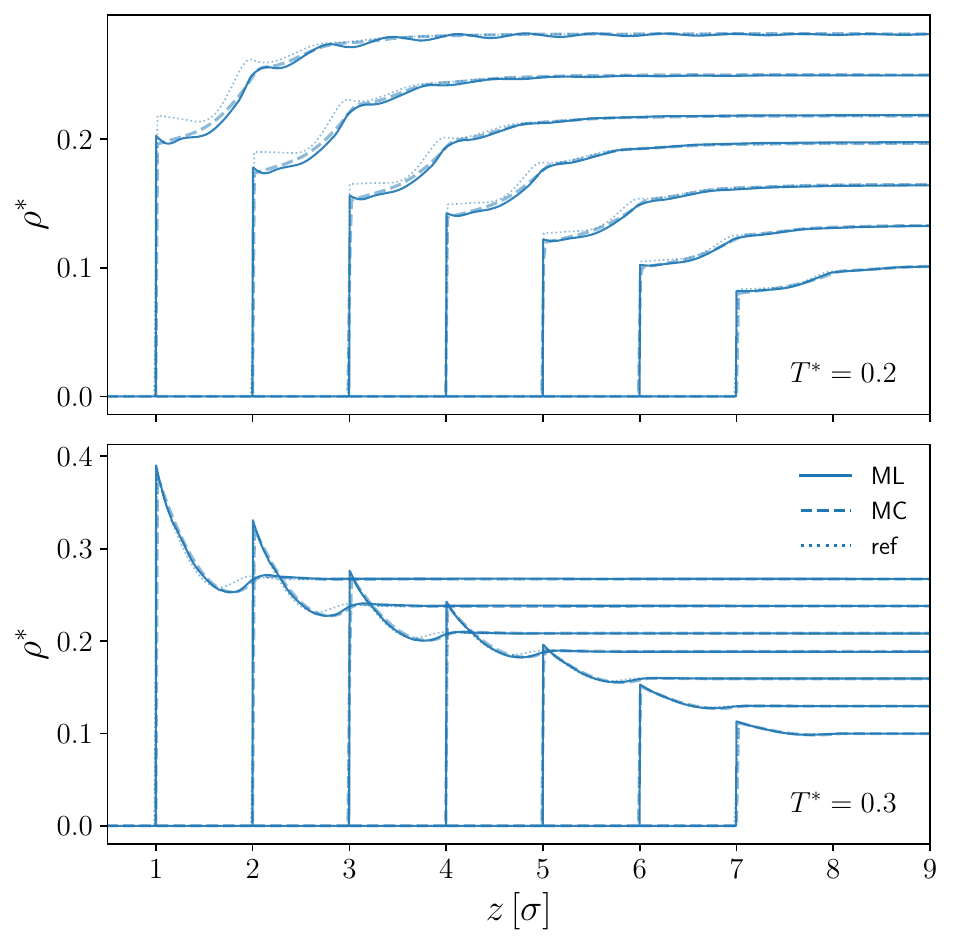}
  \caption{Self-consistent density profiles from the trained ML-functional with the mean-field density correction  ($M^{00}$ kernel). The high-temperature regime works well for all densities, whereas for low temperatures unphysical oscillations appear at high densities.
  Bulk densities are from the set $\{0.1,\, 0.13,\, 0.16,\, 0.19,\, 0.21,\, 0.24,\, 0.27\}$, labeled with $i=0...6$.
  Profiles for different bulk densities $\rho_i$ are plotted with an offset $i\cdot \sigma$ in the $z$ direction. }
  \label{fig:rho_mf_correction}
\end{figure}

The results for the density profile in \cref{fig:kf_comparison_more} showed some shortcomings of the Stopper--Wu reference functional especially at the lowest temperature $T^*=0.2$. We tackle these 
by allowing for a mean-field density correction through learning the kernel moment $M^{00}$. The self-consistent
results for the orientational moment and density profiles are shown in \cref{fig:ori_mf_correction,fig:rho_mf_correction}, again for the lowest temperature $T^*=0.2$ and the highest temperature $T^*=0.3$ in the dataset.
We observe no significant change in the description of the orientational moments (compare figs.~\ref{fig:ori_mf_correction} and \ref{fig:ori_fixed_ref}). For $T^*=0.3$ the already good agreement with simulation data has further improved slightly, but the qualitative deficits for $T^*=0.2$ especially at low densities remain. For low temperatures and low densities one expects a very high bonding fraction with especially strong correlations in the orientations of the bonded particles. Although the high bonding fraction is captured more or less correctly in the Stopper--Wu reference functional, the associated orientational correlations are captured insufficiently with the orientational mean-field ansatz which presumably misses higher order correlations (e.g. an associated direct correlation function of third order is zero).   

\begin{figure}[h]
  \centering
  \includegraphics[width=\columnwidth]{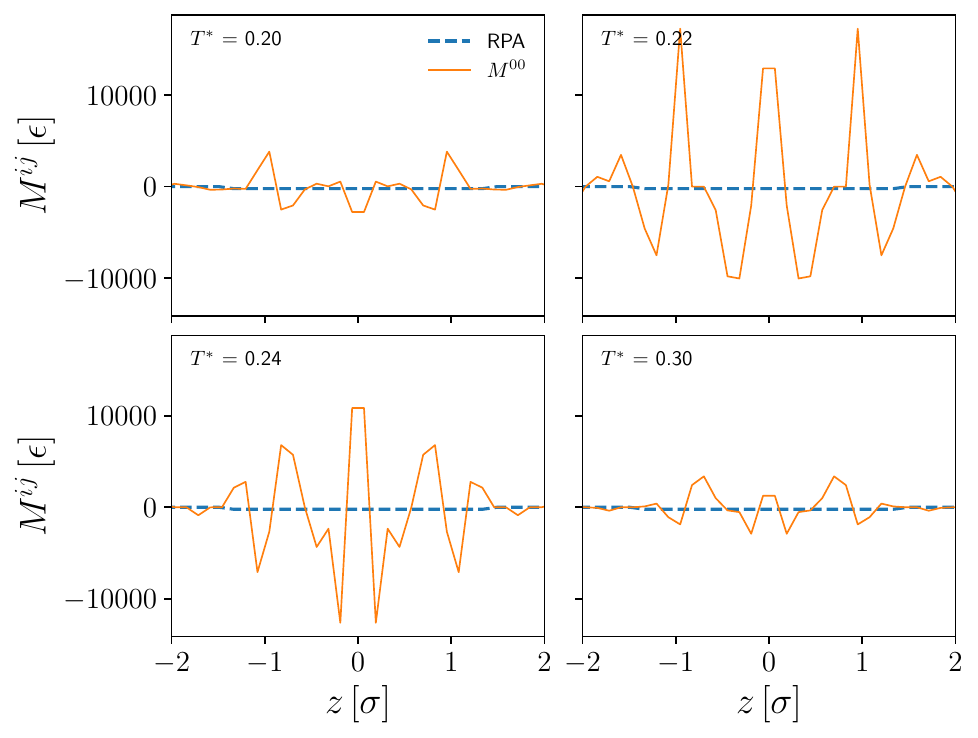}
  \caption{Learned mean-field density correction (kernel moment $M^{00}$ in units of $\epsilon$) for different temperatures together with the RPA result. Here the structure strongly varies with the temperature. The dashed line (RPA) has a magnitude of $210$ in the center region.}
  \label{fig:correction_isotherm_rho}
\end{figure}

The density profiles for the highest temperature $T^*=0.3$ (\cref{fig:rho_mf_correction}) are fitted quite well with the mean-field correction in $\Delta \FE^\text{mf}_{\text{ex,iso}}$. For the lowest temperature $T^*=0.2$ we observe an improved wall contact density which is however accompanied by small, unphysical oscillations of the density profile with a wavelength of $\approx 0.7 \sigma$. The reason for these oscillations can be found in the $z$--dependence of the reduced kernel moment $M^{00}$ shown in \cref{fig:correction_isotherm_rho} (together with the corresponding RPA result). All learned  $M^{00}$ show oscillations with similar wavelengths and an amplitude much larger than the RPA moment. Note, however, that $a:=\int z \odif{z} M^{00}(z)$ is small: for bulk systems the correction in the free energy density is $\Delta f = (a/2)\rho^2$ which likewise leads to a correction in the bulk pressure $\Delta p = (a/2)\rho^2$, and the latter determines the contact density $\rho_c$ at the wall. Thus the ML fit (through $\rho_c$) constrains well the integral of $M^{00}$ (i.e. the equation of state) but not the shape.      

The resulting kernel moments are slightly different than those obtained in \cref{sec:sub_fixed}. This is, for one, due the fact that another density profile is used (owing to the correction term) and also, that an additional potential enters into the $L^1$ regularization, thus adding a further coupling between the regularization and the resulting density profiles.
The notable influence of the $L^1$ regularization on the kernel moments and the resulting orientational moment profiles is further discussed in \cref{appendix:reg}.

\begin{figure}[t]
  \centering
  \includegraphics[width=\columnwidth]{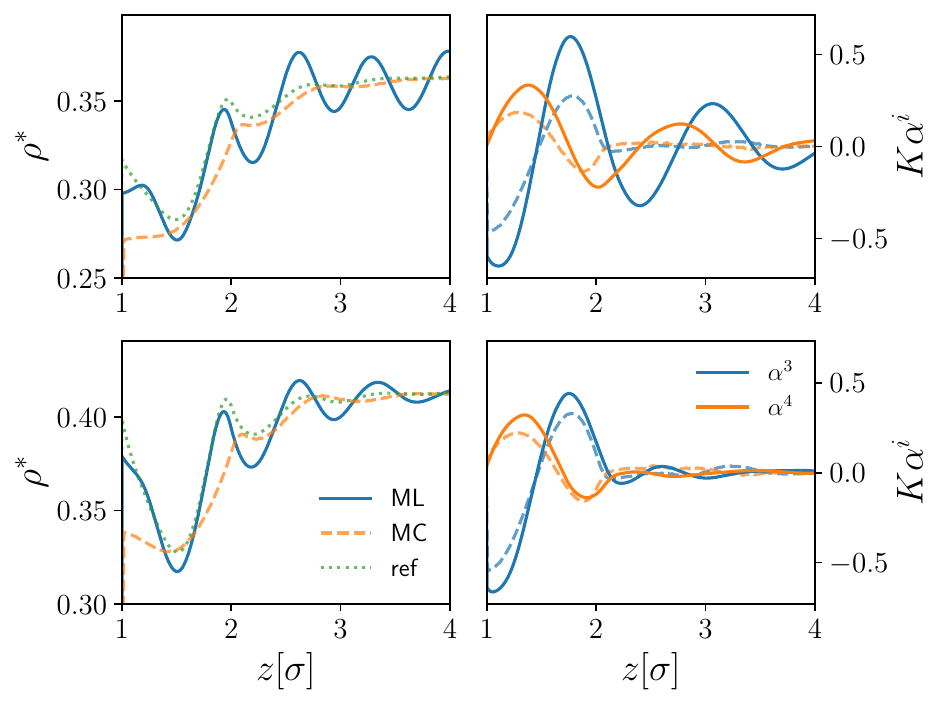}
  \caption{Learned functional for the $T^*=0.20$ isotherm evaluated at densities
  outside of the trainig data set. Unphysical long-range oscillations of wavelength $\lambda \approx 0.7$ are clearly visible.}
  \label{fig:extrapol_low}
\end{figure}

\begin{figure}[t]
  \centering
  \includegraphics[width=\columnwidth]{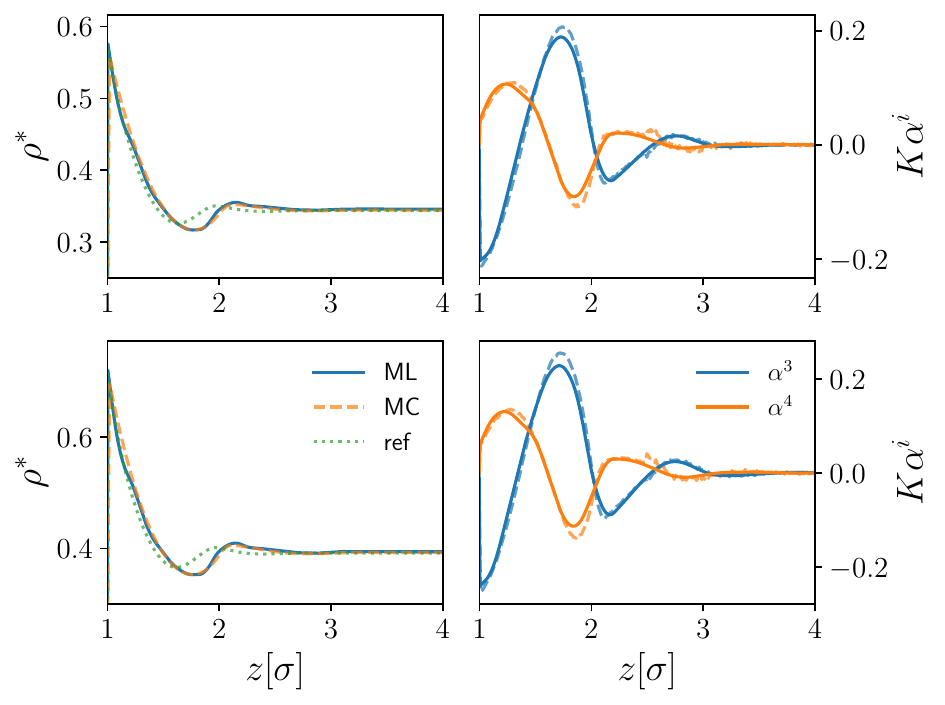}
  \caption{Extrapolation of the learned functional at the $T^*=0.30$ isotherm.
   Both densities and orientation agree reasonably well.}
  \label{fig:extrapol_high}
\end{figure}

\subsection{Extrapolation to higher densities}

In order to assess the quality of the learned functional (with mean-field correction of the reference functional) we compute density and orientation profiles at state points outside of the original training set (validation set) and compare to simulations. 
For every isotherm (trained on reduced densities between $0.10$ and $0.27$), we extrapolate now to $0.35$ and $0.40$. The resulting density and orientation profiles are shown in \cref{fig:extrapol_low,fig:extrapol_high}. The high-temperature case agrees reasonably well with simulations and confirms the adequacy of a mean-field treatment of the  orientational part. For the low temperature 0.20, we observe unphysical oscillations in the density profile, and the profiles of the orientational moments are described only semi-quantitatively. Overall, this underpins all observations that we already made in fitting the training set, especially the shortcomings of the mean-field approximation.
Also the oscillations in the density profile are of the same origin as discussed before. We checked additionally that the
correction in the free energy density $\Delta f$ and the bulk pressure $\Delta p$ did not bring the system inside the coexistence region, i.e. there is no van der Waals loop in the pressure. A quantitative investigation of the unphysical density oscillations would require an investigation of the complex poles of the direct correlation function of the learned functional, similar as in~\cite{grodon2004decay}.

\section{Summary and outlook}
\label{sec:summary}

We have investigated the full orientational structure of a 4-patch associating fluid with tetrahedral symmetry between hard walls for supercritical conditions at different temperatures and densities, with the lowest temperature  fairly close to the critical point. Results from grand canonical Monte Carlo simulations show a rather strong anisotropy close to the walls. For a description of this system using classical density functional theory, we propose a splitting of the excess free energy functional into a reference part, depending on the orientationally averaged density $\rho(\mathbf{r})$, and an orientation dependent part, modelled by a general mean-field ansatz. For the reference part, the Stopper--Wu functional of ref.~\cite{stopper2018bulk} is used which takes into account the patch connectivity as in Wertheim's perturbation theory for associating fluids. The Stopper--Wu functional describes the phase diagram and averaged density profiles semi-quantitatively with the strongest deviations for low temperatures close to the critical point, thus motivating the inclusion of the orientational part of the free energy functional. Orientational distributions are expanded in Wigner D-matrices as base functions and a projection scheme for particle symmetry-adapted base functions has been laid out. 
We have investigated two mean-field kernels for the orientational free energy: (i) from the random phase approximation with the kernel calculated by Monte Carlo integration of the Kern-Frenkel interparticle potential and (ii) temperature-dependent kernels from machine learning, trained on the simulation data.

RPA results for the orientational profiles are too small by more than an order of magnitude and thus RPA is inadequate. Using ML, we succeeded to train a functional which could be stably minimized. For the highest temperature investigated
($T^*$=0.3), the orientational profiles compare very well to simulation, for the lowest temperature $T^*$=0.2 substantial deviations are seen especially in the low--density region, signalling the break-down of the mean-field ansatz (the critical temperature $T^*_c \approx 0.17$). The ML mean-field kernel is explicitly temperature-dependent, is much stronger than the RPA kernel but the qualitative shape of the kernel moments is similar. We also investigated corrections to the average density with this mean-field ansatz which also works well for higher temperatures but results in unphysical density oscillations at lower temperatures.

The present work should be considered as a starting point for further investigations of ML-techniques to learn functionals for anisotropic systems. To the best of our knowledge, this is the first work where ML is used to learn a largely unknown functional (here the orientational part of the functional for the Kern-Frenke fluid). There are clearly some differences to isotropic fluids which had been studied before. First, there is considerably more analytic ''overhead'' work owing to the orientational expansions. Secondly, the restriction to training data at flat walls is insufficient for learning the full mean-field kernel (only a reduced set of moments is addressed). This is different to the LJ fluid \cite{cats2021}, say.     
Therefore, future work should address the problem of a suitable choice of training geometries and of lifting the mean-field restriction. The latter can be achieved by simply allowing more general analytic forms with unknown parameters \cite{cats2021, yang1994density}, by learning an analytic form itself \cite{shang2020FEQL} or by learning a ''black box'' representation of the map between the full orientation-dependent density and the first-order direct correlation function \cite{sammuller2023neural}.  

A possibly rich application background for ML-DFT for anisotropic fluids is the case of DFT for water and solvation of molecules in water or ``industrial'' fluids such as CO2 . This route has been pursued by Borgis and coworkers in the past years~\cite{borgis2021accurate, mohamed2023exact} and an open point is the full orientation-dependent of the ''bridge'' functional which comprises all terms in the functional beyond the second-order terms in a functional expansion around a bulk reference state.

\textbf{Acknowledgment:} We gratefully acknowledge funding by the Deutsche Forschungsgemeinschaft (DFG, German Research Foundation) under Germany’s Excellence Strategy – EXC number 2064/1 – Project
 number 390727645.

\appendix

\section{Results for other isotherms}

\begin{figure}[h]
  \centering
  \includegraphics[width=\columnwidth]{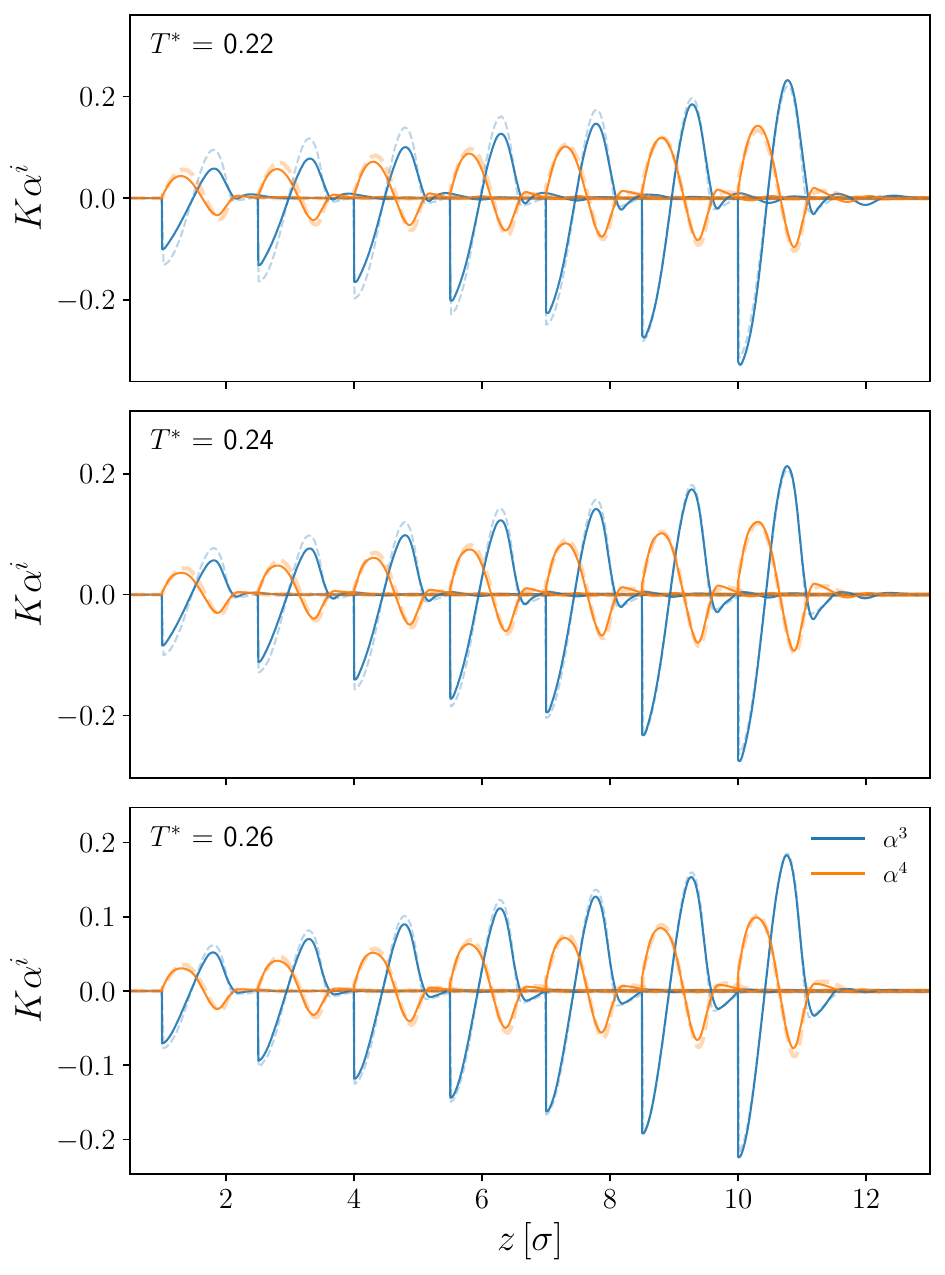}
  \caption{Self-consistent solution of the orientational moments from ML-$\FE_{\text{ex,or}}$ and ML-$\Delta \FE^\text{mf}_{\text{ex,iso}}$ trained at different temperatures. The general trend of improved predictions at higher temperatures and densities is clearly visible.
  Dashed lines close to the full lines  are the respective simulation results.}
  \label{fig:alphas_rest}
\end{figure}

In the main text we limited ourselves to the highest and lowest temperature cases when discussing the results of the self-consistent orientation profiles. We fill the gaps 
by showing also the temperatures in between, i.e. \numlist{0.22;0.24;0.26}, see \cref{fig:alphas_rest}. When lowering the temperature, the accuracy for the orientational profiles at low densities gradually worsens. 

\section{Effect of the regularization}
\label{appendix:reg}
\begin{figure}[h]
  \centering
  \includegraphics[width=\columnwidth]{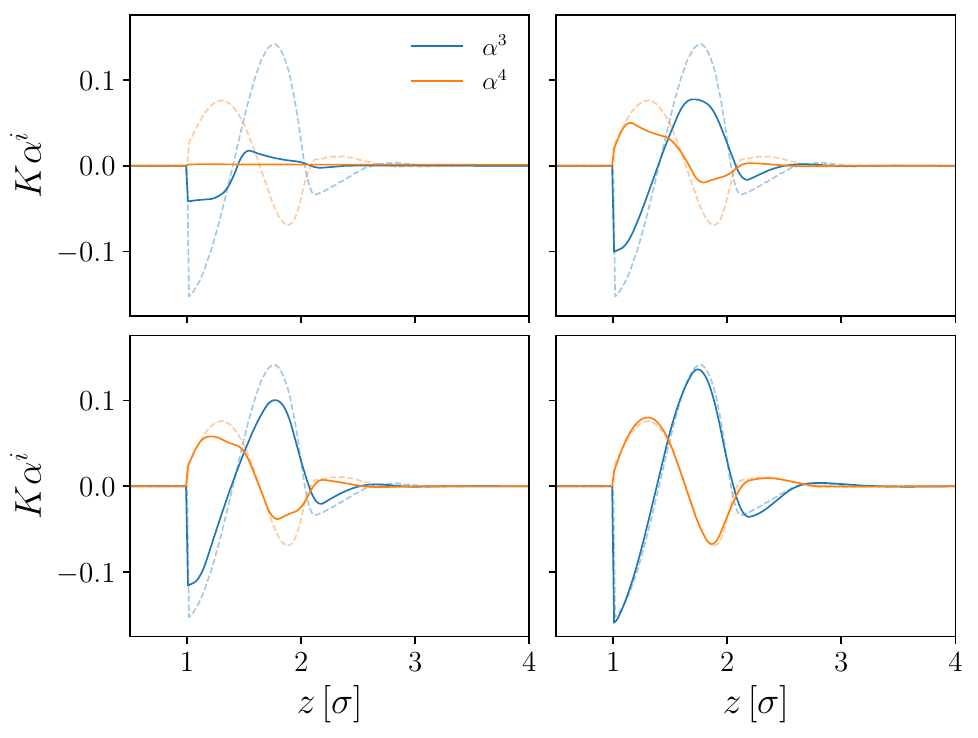}
  \caption{Self-consistent solution of the orientational moments close to the hard wall for $T^*=0.3$ and bulk density $\rho^*=0.27$, for varying $L^1$ regularization strength. The corresponding kernels are shown in \cref{fig:kernels_lamb}. The individual values for $\lambda$ were \numlist{3e-3;1e-3;5e-4; 5e-5}.
  Dashed lines are the respective simulation results.}
  \label{fig:decreasing_lamb}
\end{figure}

\begin{figure}[h]
  \centering
  \includegraphics[width=\columnwidth]{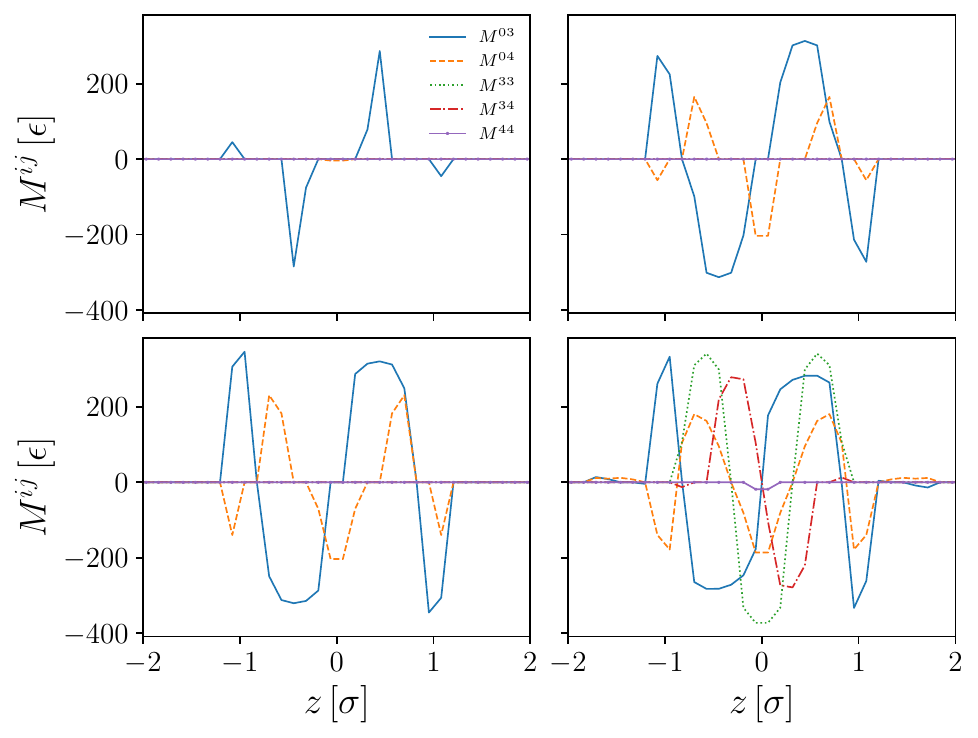}
  \caption{Learned reduced mean-field kernel moments $M^{ij}$  (in units of $\epsilon$) at one isotherm ($T^* = $ \num{0.30}) but with varying $L^1$ regularization. The individual values of $\lambda$ and the resulting profiles are shown in \cref{fig:decreasing_lamb}}
  \label{fig:kernels_lamb}
\end{figure}

As argued before, a
$L^1$ regularization is needed for  mean-field kernels that are as simple as possible and also needed to prevent the system from "fitting to the noise". The best value must then be inferred
during a hyperparameter optimization where one needs to find a compromise between simplicity and quality of prediction. 
For a too weak regularization, mean-field kernels become very noisy
making it hard  to find a sensible interpretation for them in the physical sense (i.e. attraction and repulsion). On the other hand, when the regularizer is too strong, we have the chance to look at "leading order" effects, because then the system selects features with the best signal to cost ratio. Thanks to the training procedure whcich always produces kernels that are self-consistent, we can examine solutions for the orientational moments produced by these over-regularized potentials. They are shown in \cref{fig:decreasing_lamb} together with the corresponding kernels in \cref{fig:kernels_lamb}. The pictures show training results at the same state point but with decreasing regularization strength parameter $\lambda$. The third and fourth parameter values in particular show how the first two kernels are enough to reproduce the correct orientations up to a rather good precision, but that for more accurate results the other interaction potentials need to be included (note that between the two pictures, the kernels $M^{03}$ and $M^{04}$ do not change much).

\clearpage

\bibliography{mybibliography} 

\providecommand{\latin}[1]{#1}
\makeatletter
\providecommand{\doi}
  {\begingroup\let\do\@makeother\dospecials
  \catcode`\{=1 \catcode`\}=2 \doi@aux}
\providecommand{\doi@aux}[1]{\endgroup\texttt{#1}}
\makeatother
\providecommand*\mcitethebibliography{\thebibliography}
\csname @ifundefined\endcsname{endmcitethebibliography}
  {\let\endmcitethebibliography\endthebibliography}{}
\begin{mcitethebibliography}{43}
\providecommand*\natexlab[1]{#1}
\providecommand*\mciteSetBstSublistMode[1]{}
\providecommand*\mciteSetBstMaxWidthForm[2]{}
\providecommand*\mciteBstWouldAddEndPuncttrue
  {\def\EndOfBibitem{\unskip.}}
\providecommand*\mciteBstWouldAddEndPunctfalse
  {\let\EndOfBibitem\relax}
\providecommand*\mciteSetBstMidEndSepPunct[3]{}
\providecommand*\mciteSetBstSublistLabelBeginEnd[3]{}
\providecommand*\EndOfBibitem{}
\mciteSetBstSublistMode{f}
\mciteSetBstMaxWidthForm{subitem}{(\alph{mcitesubitemcount})}
\mciteSetBstSublistLabelBeginEnd
  {\mcitemaxwidthsubitemform\space}
  {\relax}
  {\relax}

\bibitem[Bianchi \latin{et~al.}(2006)Bianchi, Largo, Tartaglia, Zaccarelli, and
  Sciortino]{bianchi_phase_2006}
Bianchi,~E.; Largo,~J.; Tartaglia,~P.; Zaccarelli,~E.; Sciortino,~F. Phase
  {Diagram} of {Patchy} {Colloids}: {Towards} {Empty} {Liquids}. \emph{Physical
  Review Letters} \textbf{2006}, \emph{97}, 168301, Publisher: American
  Physical Society\relax
\mciteBstWouldAddEndPuncttrue
\mciteSetBstMidEndSepPunct{\mcitedefaultmidpunct}
{\mcitedefaultendpunct}{\mcitedefaultseppunct}\relax
\EndOfBibitem
\bibitem[Heras \latin{et~al.}(2011)Heras, Tavares, and
  da~Gama]{heras_phase_2011}
Heras,~D. d.~l.; Tavares,~J.~M.; da~Gama,~M. M.~T. Phase diagrams of binary
  mixtures of patchy colloids with distinct numbers and types of patches: {The}
  empty fluid regime. \emph{The Journal of Chemical Physics} \textbf{2011},
  \emph{134}, 104904\relax
\mciteBstWouldAddEndPuncttrue
\mciteSetBstMidEndSepPunct{\mcitedefaultmidpunct}
{\mcitedefaultendpunct}{\mcitedefaultseppunct}\relax
\EndOfBibitem
\bibitem[Smallenburg and Sciortino(2013)Smallenburg, and
  Sciortino]{smallenburg_liquids_2013}
Smallenburg,~F.; Sciortino,~F. Liquids more stable than crystals in particles
  with limited valence and flexible bonds. \emph{Nature Physics} \textbf{2013},
  \emph{9}, 554--558, Number: 9 Publisher: Nature Publishing Group\relax
\mciteBstWouldAddEndPuncttrue
\mciteSetBstMidEndSepPunct{\mcitedefaultmidpunct}
{\mcitedefaultendpunct}{\mcitedefaultseppunct}\relax
\EndOfBibitem
\bibitem[Sciortino and Zaccarelli(2017)Sciortino, and
  Zaccarelli]{sciortino_equilibrium_2017}
Sciortino,~F.; Zaccarelli,~E. Equilibrium gels of limited valence colloids.
  \emph{Current Opinion in Colloid \& Interface Science} \textbf{2017},
  \emph{30}, 90--96\relax
\mciteBstWouldAddEndPuncttrue
\mciteSetBstMidEndSepPunct{\mcitedefaultmidpunct}
{\mcitedefaultendpunct}{\mcitedefaultseppunct}\relax
\EndOfBibitem
\bibitem[Kern and Frenkel(2003)Kern, and Frenkel]{kern_frenkel2003}
Kern,~N.; Frenkel,~D. Fluid--fluid coexistence in colloidal systems with
  short-ranged strongly directional attraction. \emph{The Journal of chemical
  physics} \textbf{2003}, \emph{118}, 9882--9889\relax
\mciteBstWouldAddEndPuncttrue
\mciteSetBstMidEndSepPunct{\mcitedefaultmidpunct}
{\mcitedefaultendpunct}{\mcitedefaultseppunct}\relax
\EndOfBibitem
\bibitem[Foffi and Sciortino(2007)Foffi, and Sciortino]{foffi2007possibility}
Foffi,~G.; Sciortino,~F. On the possibility of extending the noro- frenkel
  generalized law of correspondent states to nonisotropic patchy interactions.
  \emph{The Journal of Physical Chemistry B} \textbf{2007}, \emph{111},
  9702--9705\relax
\mciteBstWouldAddEndPuncttrue
\mciteSetBstMidEndSepPunct{\mcitedefaultmidpunct}
{\mcitedefaultendpunct}{\mcitedefaultseppunct}\relax
\EndOfBibitem
\bibitem[Romano \latin{et~al.}(2010)Romano, Sanz, and
  Sciortino]{romano2010phase}
Romano,~F.; Sanz,~E.; Sciortino,~F. Phase diagram of a tetrahedral patchy
  particle model for different interaction ranges. \emph{The Journal of
  Chemical Physics} \textbf{2010}, \emph{132}, 184501\relax
\mciteBstWouldAddEndPuncttrue
\mciteSetBstMidEndSepPunct{\mcitedefaultmidpunct}
{\mcitedefaultendpunct}{\mcitedefaultseppunct}\relax
\EndOfBibitem
\bibitem[Lutsko(2010)]{lutsko2010recent}
Lutsko,~J.~F. Recent developments in classical density functional theory.
  \emph{Advances in chemical physics} \textbf{2010}, \emph{144}, 1\relax
\mciteBstWouldAddEndPuncttrue
\mciteSetBstMidEndSepPunct{\mcitedefaultmidpunct}
{\mcitedefaultendpunct}{\mcitedefaultseppunct}\relax
\EndOfBibitem
\bibitem[Evans(2009)]{evans2009density}
Evans,~R. Density functional theory for inhomogeneous fluids I: Simple fluids
  in equilibrium. \emph{Lectures at 3rd Warsaw School of Statistical Physics,
  Kazimierz Dolny} \textbf{2009}, \emph{27}\relax
\mciteBstWouldAddEndPuncttrue
\mciteSetBstMidEndSepPunct{\mcitedefaultmidpunct}
{\mcitedefaultendpunct}{\mcitedefaultseppunct}\relax
\EndOfBibitem
\bibitem[Roth(2010)]{roth2010fundamental}
Roth,~R. Fundamental measure theory for hard-sphere mixtures: a review.
  \emph{Journal of Physics: Condensed Matter} \textbf{2010}, \emph{22},
  063102\relax
\mciteBstWouldAddEndPuncttrue
\mciteSetBstMidEndSepPunct{\mcitedefaultmidpunct}
{\mcitedefaultendpunct}{\mcitedefaultseppunct}\relax
\EndOfBibitem
\bibitem[Yu and Wu(2002)Yu, and Wu]{yuwu2002}
Yu,~Y.-X.; Wu,~J. A fundamental-measure theory for inhomogeneous associating
  fluids. \emph{The Journal of chemical physics} \textbf{2002}, \emph{116},
  7094--7103\relax
\mciteBstWouldAddEndPuncttrue
\mciteSetBstMidEndSepPunct{\mcitedefaultmidpunct}
{\mcitedefaultendpunct}{\mcitedefaultseppunct}\relax
\EndOfBibitem
\bibitem[Stopper \latin{et~al.}(2018)Stopper, Hirschmann, Oettel, and
  Roth]{stopper2018bulk}
Stopper,~D.; Hirschmann,~F.; Oettel,~M.; Roth,~R. Bulk structural information
  from density functionals for patchy particles. \emph{The Journal of Chemical
  Physics} \textbf{2018}, \emph{149}, 224503\relax
\mciteBstWouldAddEndPuncttrue
\mciteSetBstMidEndSepPunct{\mcitedefaultmidpunct}
{\mcitedefaultendpunct}{\mcitedefaultseppunct}\relax
\EndOfBibitem
\bibitem[Shang-Chun and Oettel(2019)Shang-Chun, and Oettel]{shang2019classical}
Shang-Chun,~L.; Oettel,~M. A classical density functional from machine learning
  and a convolutional neural network. \emph{SciPost Physics} \textbf{2019},
  \emph{6}, 025\relax
\mciteBstWouldAddEndPuncttrue
\mciteSetBstMidEndSepPunct{\mcitedefaultmidpunct}
{\mcitedefaultendpunct}{\mcitedefaultseppunct}\relax
\EndOfBibitem
\bibitem[Lin \latin{et~al.}(2020)Lin, Martius, and Oettel]{shang2020FEQL}
Lin,~S.-C.; Martius,~G.; Oettel,~M. Analytical classical density functionals
  from an equation learning network. \emph{The Journal of Chemical Physics}
  \textbf{2020}, \emph{152}, 021102\relax
\mciteBstWouldAddEndPuncttrue
\mciteSetBstMidEndSepPunct{\mcitedefaultmidpunct}
{\mcitedefaultendpunct}{\mcitedefaultseppunct}\relax
\EndOfBibitem
\bibitem[Yatsyshina \latin{et~al.}(2020)Yatsyshina, Kalliadasisb, and
  Duncana]{yatsyshin2021}
Yatsyshina,~P.; Kalliadasisb,~S.; Duncana,~A.~B. Data Driven Classical Density
  Functional Theory: A case for Physics Informed Learning. \emph{arXiv preprint
  arXiv:2010.03374} \textbf{2020}, \relax
\mciteBstWouldAddEndPunctfalse
\mciteSetBstMidEndSepPunct{\mcitedefaultmidpunct}
{}{\mcitedefaultseppunct}\relax
\EndOfBibitem
\bibitem[Zhang \latin{et~al.}(2022)Zhang, Wu, Xing, Zhang, Wang, Feng, Zhu, Lu,
  and Mu]{wu2022}
Zhang,~T.; Wu,~C.; Xing,~Z.; Zhang,~J.; Wang,~S.; Feng,~X.; Zhu,~J.; Lu,~X.;
  Mu,~L. Machine Learning Prediction of Photocatalytic Lignin Cleavage of CC
  Bonds based on Density Functional Theory. \emph{Materials Today
  Sustainability} \textbf{2022}, 100256\relax
\mciteBstWouldAddEndPuncttrue
\mciteSetBstMidEndSepPunct{\mcitedefaultmidpunct}
{\mcitedefaultendpunct}{\mcitedefaultseppunct}\relax
\EndOfBibitem
\bibitem[Cats \latin{et~al.}(2021)Cats, Kuipers, De~Wind, Van~Damme, Coli,
  Dijkstra, and Van~Roij]{cats2021}
Cats,~P.; Kuipers,~S.; De~Wind,~S.; Van~Damme,~R.; Coli,~G.~M.; Dijkstra,~M.;
  Van~Roij,~R. Machine-learning free-energy functionals using density profiles
  from simulations. \emph{APL Materials} \textbf{2021}, \emph{9}, 031109\relax
\mciteBstWouldAddEndPuncttrue
\mciteSetBstMidEndSepPunct{\mcitedefaultmidpunct}
{\mcitedefaultendpunct}{\mcitedefaultseppunct}\relax
\EndOfBibitem
\bibitem[Samm{\"u}ller \latin{et~al.}(2023)Samm{\"u}ller, Hermann, Heras, and
  Schmidt]{sammuller2023neural}
Samm{\"u}ller,~F.; Hermann,~S.; Heras,~D. d.~l.; Schmidt,~M. Neural functional
  theory for inhomogeneous fluids: Fundamentals and applications. \emph{arXiv
  preprint arXiv:2307.04539} \textbf{2023}, \relax
\mciteBstWouldAddEndPunctfalse
\mciteSetBstMidEndSepPunct{\mcitedefaultmidpunct}
{}{\mcitedefaultseppunct}\relax
\EndOfBibitem
\bibitem[Snyder \latin{et~al.}(2012)Snyder, Rupp, Hansen, M{\"u}ller, and
  Burke]{burke2012}
Snyder,~J.~C.; Rupp,~M.; Hansen,~K.; M{\"u}ller,~K.-R.; Burke,~K. Finding
  density functionals with machine learning. \emph{Physical review letters}
  \textbf{2012}, \emph{108}, 253002\relax
\mciteBstWouldAddEndPuncttrue
\mciteSetBstMidEndSepPunct{\mcitedefaultmidpunct}
{\mcitedefaultendpunct}{\mcitedefaultseppunct}\relax
\EndOfBibitem
\bibitem[Ma \latin{et~al.}(2022)Ma, Narayanaswamy, Riley, and Li]{lili2022}
Ma,~H.; Narayanaswamy,~A.; Riley,~P.; Li,~L. Evolving symbolic density
  functionals. \emph{arXiv preprint arXiv:2203.02540} \textbf{2022}, \relax
\mciteBstWouldAddEndPunctfalse
\mciteSetBstMidEndSepPunct{\mcitedefaultmidpunct}
{}{\mcitedefaultseppunct}\relax
\EndOfBibitem
\bibitem[Li \latin{et~al.}(2021)Li, Hoyer, Pederson, Sun, Cubuk, Riley, and
  Burke]{lili2021}
Li,~L.; Hoyer,~S.; Pederson,~R.; Sun,~R.; Cubuk,~E.~D.; Riley,~P.; Burke,~K.
  Kohn-Sham Equations as Regularizer: Building Prior Knowledge into
  Machine-Learned Physics. \emph{Phys. Rev. Lett.} \textbf{2021}, \emph{126},
  036401\relax
\mciteBstWouldAddEndPuncttrue
\mciteSetBstMidEndSepPunct{\mcitedefaultmidpunct}
{\mcitedefaultendpunct}{\mcitedefaultseppunct}\relax
\EndOfBibitem
\bibitem[Cattes \latin{et~al.}(2016)Cattes, Gubbins, and Schoen]{cattes2016}
Cattes,~S.~M.; Gubbins,~K.~E.; Schoen,~M. Mean-field density functional theory
  of a nanoconfined classical, three-dimensional Heisenberg fluid. I. The role
  of molecular anchoring. \emph{The Journal of Chemical Physics} \textbf{2016},
  \emph{144}, 194704\relax
\mciteBstWouldAddEndPuncttrue
\mciteSetBstMidEndSepPunct{\mcitedefaultmidpunct}
{\mcitedefaultendpunct}{\mcitedefaultseppunct}\relax
\EndOfBibitem
\bibitem[Wandrei \latin{et~al.}(2018)Wandrei, Roth, and
  Schoen]{wandrei2018mean}
Wandrei,~S.~M.; Roth,~R.; Schoen,~M. Mean-field density functional theory of a
  nanoconfined classical, three-dimensional Heisenberg fluid. II. The interplay
  between molecular packing and orientational order. \emph{The Journal of
  Chemical Physics} \textbf{2018}, \emph{149}, 054704\relax
\mciteBstWouldAddEndPuncttrue
\mciteSetBstMidEndSepPunct{\mcitedefaultmidpunct}
{\mcitedefaultendpunct}{\mcitedefaultseppunct}\relax
\EndOfBibitem
\bibitem[Teixeira and Sciortino(2019)Teixeira, and
  Sciortino]{teixeira2019patchy}
Teixeira,~P.; Sciortino,~F. Patchy particles at a hard wall:
  Orientation-dependent bonding. \emph{The Journal of Chemical Physics}
  \textbf{2019}, \emph{151}, 174903\relax
\mciteBstWouldAddEndPuncttrue
\mciteSetBstMidEndSepPunct{\mcitedefaultmidpunct}
{\mcitedefaultendpunct}{\mcitedefaultseppunct}\relax
\EndOfBibitem
\bibitem[Blum and Torruella(1972)Blum, and Torruella]{blum1972invariant}
Blum,~L.; Torruella,~A. Invariant Expansion for Two-Body Correlations:
  Thermodynamic Functions, Scattering, and the Ornstein—Zernike Equation.
  \emph{The Journal of Chemical Physics} \textbf{1972}, \emph{56},
  303--310\relax
\mciteBstWouldAddEndPuncttrue
\mciteSetBstMidEndSepPunct{\mcitedefaultmidpunct}
{\mcitedefaultendpunct}{\mcitedefaultseppunct}\relax
\EndOfBibitem
\bibitem[Ding \latin{et~al.}(2017)Ding, Levesque, Borgis, and
  Belloni]{ding2017efficient}
Ding,~L.; Levesque,~M.; Borgis,~D.; Belloni,~L. Efficient molecular density
  functional theory using generalized spherical harmonics expansions. \emph{The
  Journal of Chemical Physics} \textbf{2017}, \emph{147}, 094107\relax
\mciteBstWouldAddEndPuncttrue
\mciteSetBstMidEndSepPunct{\mcitedefaultmidpunct}
{\mcitedefaultendpunct}{\mcitedefaultseppunct}\relax
\EndOfBibitem
\bibitem[Belloni(2017)]{belloni2017exact}
Belloni,~L. Exact molecular direct, cavity, and bridge functions in water
  system. \emph{The Journal of Chemical Physics} \textbf{2017},
  \emph{147}\relax
\mciteBstWouldAddEndPuncttrue
\mciteSetBstMidEndSepPunct{\mcitedefaultmidpunct}
{\mcitedefaultendpunct}{\mcitedefaultseppunct}\relax
\EndOfBibitem
\bibitem[Jeanmairet \latin{et~al.}(2013)Jeanmairet, Levesque, Vuilleumier, and
  Borgis]{jeanmairet2013molecular}
Jeanmairet,~G.; Levesque,~M.; Vuilleumier,~R.; Borgis,~D. Molecular density
  functional theory of water. \emph{The Journal of Physical Chemistry Letters}
  \textbf{2013}, \emph{4}, 619--624\relax
\mciteBstWouldAddEndPuncttrue
\mciteSetBstMidEndSepPunct{\mcitedefaultmidpunct}
{\mcitedefaultendpunct}{\mcitedefaultseppunct}\relax
\EndOfBibitem
\bibitem[Borgis \latin{et~al.}(2021)Borgis, Luukkonen, Belloni, and
  Jeanmairet]{borgis2021accurate}
Borgis,~D.; Luukkonen,~S.; Belloni,~L.; Jeanmairet,~G. Accurate prediction of
  hydration free energies and solvation structures using molecular density
  functional theory with a simple bridge functional. \emph{The Journal of
  Chemical Physics} \textbf{2021}, \emph{155}\relax
\mciteBstWouldAddEndPuncttrue
\mciteSetBstMidEndSepPunct{\mcitedefaultmidpunct}
{\mcitedefaultendpunct}{\mcitedefaultseppunct}\relax
\EndOfBibitem
\bibitem[Yang \latin{et~al.}(1994)Yang, Sullivan, and Gray]{yang1994density}
Yang,~B.; Sullivan,~D.; Gray,~C. Density-functional theory of the water
  liquid-vapour interface: II. \emph{Journal of Physics: Condensed Matter}
  \textbf{1994}, \emph{6}, 4823\relax
\mciteBstWouldAddEndPuncttrue
\mciteSetBstMidEndSepPunct{\mcitedefaultmidpunct}
{\mcitedefaultendpunct}{\mcitedefaultseppunct}\relax
\EndOfBibitem
\bibitem[Steele(1980)]{steele1980symmetry}
Steele,~W.~A. Symmetry constraints on the configurational properties of
  non-linear molecules: Tetrahedra. \emph{Molecular Physics} \textbf{1980},
  \emph{39}, 1411--1422\relax
\mciteBstWouldAddEndPuncttrue
\mciteSetBstMidEndSepPunct{\mcitedefaultmidpunct}
{\mcitedefaultendpunct}{\mcitedefaultseppunct}\relax
\EndOfBibitem
\bibitem[Rovigatti \latin{et~al.}(2018)Rovigatti, Russo, and
  Romano]{rovigatti2018simulate}
Rovigatti,~L.; Russo,~J.; Romano,~F. How to simulate patchy particles.
  \emph{The European Physical Journal E} \textbf{2018}, \emph{41}, 1--12\relax
\mciteBstWouldAddEndPuncttrue
\mciteSetBstMidEndSepPunct{\mcitedefaultmidpunct}
{\mcitedefaultendpunct}{\mcitedefaultseppunct}\relax
\EndOfBibitem
\bibitem[Rose(1995)]{rose1995elementary}
Rose,~M.~E. \emph{Elementary theory of angular momentum}; Courier Corporation,
  1995\relax
\mciteBstWouldAddEndPuncttrue
\mciteSetBstMidEndSepPunct{\mcitedefaultmidpunct}
{\mcitedefaultendpunct}{\mcitedefaultseppunct}\relax
\EndOfBibitem
\bibitem[Gray \latin{et~al.}(1984)Gray, Gubbins, and Joslin]{gray84theory}
Gray,~C.~G.; Gubbins,~K.~E.; Joslin,~C.~G. \emph{Theory of Molecular Fluids:
  Volume 1: Fundamentals}; Oxford University Press, 1984; Vol.~10\relax
\mciteBstWouldAddEndPuncttrue
\mciteSetBstMidEndSepPunct{\mcitedefaultmidpunct}
{\mcitedefaultendpunct}{\mcitedefaultseppunct}\relax
\EndOfBibitem
\bibitem[Varshalovich \latin{et~al.}(1988)Varshalovich, Moskalev, and
  Khersonskii]{varshalovich1988quantum}
Varshalovich,~D.~A.; Moskalev,~A.~N.; Khersonskii,~V.~K. \emph{Quantum theory
  of angular momentum}; World Scientific, 1988\relax
\mciteBstWouldAddEndPuncttrue
\mciteSetBstMidEndSepPunct{\mcitedefaultmidpunct}
{\mcitedefaultendpunct}{\mcitedefaultseppunct}\relax
\EndOfBibitem
\bibitem[Wertheim(1987)]{wertheim1987thermodynamic}
Wertheim,~M. Thermodynamic perturbation theory of polymerization. \emph{The
  Journal of chemical physics} \textbf{1987}, \emph{87}, 7323--7331\relax
\mciteBstWouldAddEndPuncttrue
\mciteSetBstMidEndSepPunct{\mcitedefaultmidpunct}
{\mcitedefaultendpunct}{\mcitedefaultseppunct}\relax
\EndOfBibitem
\bibitem[Stopper(2019)]{stopper2019structure}
Stopper,~D. Structure and Dynamics of Model Fluids with Anisotropic
  Interactions. Ph.D.\ thesis, Universit{\"a}t T{\"u}bingen, 2019\relax
\mciteBstWouldAddEndPuncttrue
\mciteSetBstMidEndSepPunct{\mcitedefaultmidpunct}
{\mcitedefaultendpunct}{\mcitedefaultseppunct}\relax
\EndOfBibitem
\bibitem[Chang \latin{et~al.}(2022)Chang, Griffiths, and
  Levine]{chang2022object}
Chang,~M.; Griffiths,~T.; Levine,~S. Object representations as fixed points:
  Training iterative refinement algorithms with implicit differentiation.
  \emph{Advances in Neural Information Processing Systems} \textbf{2022},
  \emph{35}, 32694--32708\relax
\mciteBstWouldAddEndPuncttrue
\mciteSetBstMidEndSepPunct{\mcitedefaultmidpunct}
{\mcitedefaultendpunct}{\mcitedefaultseppunct}\relax
\EndOfBibitem
\bibitem[Bai \latin{et~al.}(2019)Bai, Kolter, and Koltun]{bai2019deep}
Bai,~S.; Kolter,~J.~Z.; Koltun,~V. Deep equilibrium models. \emph{Advances in
  Neural Information Processing Systems} \textbf{2019}, \emph{32}\relax
\mciteBstWouldAddEndPuncttrue
\mciteSetBstMidEndSepPunct{\mcitedefaultmidpunct}
{\mcitedefaultendpunct}{\mcitedefaultseppunct}\relax
\EndOfBibitem
\bibitem[Geng \latin{et~al.}(2021)Geng, Zhang, Bai, Wang, and
  Lin]{geng2021training}
Geng,~Z.; Zhang,~X.-Y.; Bai,~S.; Wang,~Y.; Lin,~Z. On training implicit models.
  \emph{Advances in Neural Information Processing Systems} \textbf{2021},
  \emph{34}, 24247--24260\relax
\mciteBstWouldAddEndPuncttrue
\mciteSetBstMidEndSepPunct{\mcitedefaultmidpunct}
{\mcitedefaultendpunct}{\mcitedefaultseppunct}\relax
\EndOfBibitem
\bibitem[Grodon \latin{et~al.}(2004)Grodon, Dijkstra, Evans, and
  Roth]{grodon2004decay}
Grodon,~C.; Dijkstra,~M.; Evans,~R.; Roth,~R. Decay of correlation functions in
  hard-sphere mixtures: Structural crossover. \emph{The Journal of chemical
  physics} \textbf{2004}, \emph{121}, 7869--7882\relax
\mciteBstWouldAddEndPuncttrue
\mciteSetBstMidEndSepPunct{\mcitedefaultmidpunct}
{\mcitedefaultendpunct}{\mcitedefaultseppunct}\relax
\EndOfBibitem
\bibitem[Mohamed \latin{et~al.}(2023)Mohamed, Belloni, Borgis, Ingrosso, and
  Carof]{mohamed2023exact}
Mohamed,~M.~H.; Belloni,~L.; Borgis,~D.; Ingrosso,~F.; Carof,~A. Exact direct
  correlations in the near critical region of CO2. \emph{arXiv preprint
  arXiv:2310.14667} \textbf{2023}, \relax
\mciteBstWouldAddEndPunctfalse
\mciteSetBstMidEndSepPunct{\mcitedefaultmidpunct}
{}{\mcitedefaultseppunct}\relax
\EndOfBibitem
\end{mcitethebibliography}
\end{document}